\documentclass{aastex}

\usepackage[onecolumn]{emulateapj5}
\usepackage{epsf}

\newcommand{\dm}{DIRTY\,\,}

\begin{document}

\title{The DIRTY Model II: Self-Consistent Treatment of Dust Heating
and Emission in a 3-D Radiative Transfer Code}

\author{
K. A. Misselt\altaffilmark{1,2,3}, Karl D. Gordon\altaffilmark{4},
Geoffrey C. Clayton\altaffilmark{1}, \& M. J. Wolff\altaffilmark{5}
}

\altaffiltext{1}{Department of Physics \& Astronomy, Louisiana State University, Baton
  Rouge, LA, 70803-4001} 
\altaffiltext{2}{present address: NASA/Goddard Space Flight Center,
  Code 685, Greenbelt, MD 20771}
\altaffiltext{3}{National Research Council/Resident Research Associate}
\altaffiltext{4}{Steward Observatory, University of Arizona, Tucson, AZ 85721}   
\altaffiltext{5}{Space Science Institute, 3100 Marine St., Suite A353, Boulder, CO
  80303-1058}

\begin{abstract}
  In this paper and a companion paper we present the \dm model, a Monte Carlo radiative
  transfer code, self-consistently including dust heating and emission, and accounting for
  the effects of the transient heating of small grains.  The code is completely general;
  the density structure of the dust, the number and type of heating sources, and their
  geometric configurations can be specified arbitrarily within the model space.  Source
  photons are tracked through the scattering and absorbing medium using Monte Carlo
  techniques and the effects of multiple scattering are included.  The dust scattering,
  absorbing, and emitting properties are calculated from realistic dust models derived by
  fitting observed extinction curves in Local Group galaxies including the Magellanic
  Clouds and the Milky Way.  The dust temperature and the emitted dust spectrum are
  calculated self consistently from the absorbed energy including the effects of
  temperature fluctuations in small grains.  Dust self-absorption is also accounted for,
  allowing the treatment of high optical depths, by treating photons emitted by the dust
  as an additional heating source and adopting an iterative radiative transfer scheme. As
  an illustrative case, we apply the \dm radiative transfer code to starburst galaxies
  wherein the heating sources are derived from stellar evolutionary synthesis models.
  Within the context of the starburst model, we examine the dependence of the ultraviolet
  to far-infrared spectral energy distribution, dust temperatures, and dust masses
  predicted by \dm on variations of the input parameters.
\end{abstract}

\section{Introduction}
\label{sec:intro}
Over the past two decades, studies of galaxies have become increasingly more quantitative
as powerful new instruments sensitive from the far-ultraviolet (UV) to the infrared (IR)
have become available.  With these observations, it has become clear that the presence of
dust has a significant effect on the observed properties of galaxies; the observed
spectral energy distribution (SED) is a complex convolution of the intrinsic SED of the
stellar populations with the physical properties of the absorbing and scattering medium
(dust), including its composition as well as its geometric relation to the stellar
sources.  In addition to complicating the interpretation of observations of galaxies, dust
is also an essential component in determining and modifying the physical conditions in the
interstellar medium (ISM) of galaxies, regulating star formation, and participating in the
chemical evolution of the galaxy.  The effects of dust are particularly pronounced in
galaxies undergoing active star formation, e.g., starburst galaxies.  A substantial
fraction of nearby galaxies ($\sim$30\%; Salzer et al. 1995) is made up of active star
forming galaxies and nearly all galaxies at high redshift display characteristics typical
of local star forming galaxies \citep{hetal98}.  Therefore, the ability to quantify the
effects of dust in interpreting galaxy observations has implications not only for the
study of starburst galaxies themselves, but also the formation and evolution of galaxies
over the age of the universe.

Quantifying the effects of dust in any astrophysical system is complicated by the
geometric relationship between the illumination source(s) and the dust, uncertainty in the
dust composition, and the structure of the scattering and absorbing medium.  These effects
are especially pronounced in galaxies, where a typical observing aperture may include
multiple, complex regions comprised of stars, gas, and dust mixed together in complicated
geometries and widely varying environments.  Differences in stellar populations,
metallicities, dust properties, and geometry can produce similar effects on observed
properties of galaxies.  Disentangling the effects of these various intrinsic galactic
properties requires the use of realistic models of the transfer of radiation including
both stars and dust.  Historically, the treatment of dust in radiative transfer
simulations has been necessarily simplistic.  In many cases, the dust distribution is
assumed to take the form of a foreground screen in analogy with stellar extinction
studies.  However, the extension of this simple geometry to more complicated systems
like galaxies can lead to severely erroneous results regarding the amount of dust present
and its effects on the observed SED (e.g., Witt, Thronson \& Capuano 1992 and references
therein).  Observationally, the interstellar medium in the Milky Way and external galaxies
possesses structure over a range of spatial scales, characterized by variations in density
over several orders of magnitude.  Finally, the physical characteristics of the dust
grains determine how the grains absorb, scatter, and re-emit stellar photons.  Many models
of the transfer of radiation through dusty media in galaxies have appeared in the
literature 
and although they have included all of these factors to some degree, owing to the
complexity of the problem, none treat them all simultaneously.  Some are restricted to
geometries that exhibit global symmetries \citep{err95,mh98,sgbd98,vd99,tav99}, while
others assume constant or continuously varying, homogeneous dust distributions
\citep{wtc92,err95,mh98,sgbd98,tav99} or do not fully treat the re-emission from grains in
the infrared \citep{wtc92,err95,tav99,whs99}.  

One of the seminal works in establishing the importance of geometry in radiative transfer
models of galaxies is that of \citet{wtc92}.  Subsequent work established the importance
of local structure, i.e.  clumpiness, in modeling the transport of radiation in dusty
media \citep{wg96,wg00}.  The models developed by these authors employ Monte Carlo
techniques to solve the radiative transfer equations through inhomogeneous dusty media
with arbitrary geometries.  The models are quite general, including the effects of multiple
scattering and non--isotropic scattering functions. The use of Monte Carlo techniques in
the radiative transfer problem allows the treatment of arbitrary 3-dimensional geometries
with no symmetries and can easily include a non--homogeneous, clumpy structure for the
scattering and absorbing medium. However, the models do not include the effects of dust
heating and re-emission. In this paper and a companion paper \citep{gmwc00}, we present
the \dm model which incorporates the strengths of the previous models and extends them to
include dust heating and re--emission in the IR.  The importance of the IR in understanding
galaxies can be readily seen by considering the absorption, scattering, and emission
properties of dust.  The absorption, scattering, and re-emission of photons by dust grains
occur in different wavelength regimes. Dust is very efficient at absorbing and scattering
UV/optical photons.  The energy absorbed by the dust is thermalized and re--emitted at IR
wavelengths.  As a result, a large fraction (approaching 100\% for heavily enshrouded
regions) of a galaxy's UV/Optical energy may be reprocessed and re--emitted by the dust at
IR wavelengths. Thus studies of galaxies that consider only the UV/optical wavelengths can
neglect a large fraction of the galaxies' energy budgets.  A successful model of the
transfer and emission of radiation in a galaxy must consistently reproduce the observed
SED of the galaxy from the UV to the far infrared (FIR) simultaneously.

Reprocessing of UV/optical photons into IR photons occurs through two basic mechanisms
depending both on the radiation field they are exposed to and the radius, $a$, of the dust
grain.  Large dust grains reach thermal equilibrium and emit as modified blackbodies with
an equilibrium temperature, $T_{eq}$. However, small grains (and also large molecules e.g.,
polycyclic aromatic hydrocarbons or PAH) have small heat capacities and the absorption of
even a single UV/optical photon can substantially heat the grain.  These small grains will
not reach an equilibrium temperature but will instead undergo temperature fluctuations
that lead to grain emission at temperatures well in excess of $T_{eq}$.  The inclusion of
small, thermally fluctuating grains in dust models is necessary to explain the excess of
near and mid-IR emission observed in a variety of systems, including galaxies (e.g.,
Sellgren 1984; Helou 1986; Boulanger \& P\'{e}rault 1988; Sauvage, Thuan \& Vigroux 1990).
In addition, the observation of prominent emission lines widely ascribed to PAH molecules
in the spectra of galaxies (e.g., Acosta-Pulido et al.\ 1996), indicates that a
realistic dust model should include such a component.  While many authors have elucidated
methods of calculating the emission from small, thermally fluctuating grains (e.g. Draine
\& Andersen 1985; Dwek 1986; Leger, d'Hendecourt \& D\'{e}fourneau 1989; Guhathakurta \&
Draine 1989), their inclusion in radiative transfer calculations has been limited. Our
extension of the \dm model includes large grains, small grains ($a > 100$~\AA\ and $a \le
100$~\AA, respectively; see \S\ref{sec:trans}), and PAH molecules and we treat the heating
and re-emission by each component in the appropriate regime.

In this paper, we present our model, concentrating on the dust heating and emission.
Details of the Monte Carlo calculations are presented in a companion paper \citep{gmwc00}.
In \S\ref{sec:dust_model}, we discuss the details of our dust grain model; we review the
relevant equations for determining the dust emission spectrum, describe our computational
method, and discuss the details of the computation in \S\ref{sec:physics}; results of the
model calculations are presented in \S\ref{sec:application} in the context of
applications of \dm to starburst galaxies, including a discussion of the response of the
model SED to variations in the input parameters, e.g., the dust grain model, global and
local geometries, heating sources (age, star formation rate [SFR]), size, and optical
depth; we conclude with a summary and outline some future directions in
\S\ref{sec:conclusion}.

\section{Dust Model}
\label{sec:dust_model}
In order to calculate the absorption and re--emission characteristics of a population of
dust grains, we must specify their composition, optical properties, and size distribution.
Our dust grain model consists of a mixture of carbonaceous (amorphous and graphitic
carbon) and silicate grains as well as PAH molecules. Although the exact composition of
interstellar dust is still a matter of debate and certainly varies in different
environments, we include these components in order to match several well observed
extinction and emission features in the interstellar medium of the Milky Way and other
galaxies.  The presence of silicate grains is inferred from prominent stretching and
bending mode features at $\sim 9.7~\micron$ and $\sim 18.5~\micron$ in the mid--infrared.
These features are observed in H~II regions, near young as well as evolved stars, and in
the integrated spectra of external galaxies \citep{r91,dw97}.  The well known 2175~\AA\ 
absorption feature, which is normally attributed to the presence of small carbonaceous
grains in the form of graphite (other possibilities for the carrier exist. See e.g. Draine
[1989], Duley \& Seahra [1999]), has been observed along lines of sight in our own galaxy
as well the Magellanic Clouds (Gordon \& Clayton 1998; Misselt, Clayton \& Gordon 1999)
and M~31 \citep{m31ext96}.  The presence of narrow emission features in the mid--infrared
implies a third dust component.  These features are associated with C--C (6.2 and
7.7~\micron) and C--H (3.3, 8.6, and 11.3~\micron) bending and stretching modes in
aromatic molecular structures (L\'{e}ger \& D'Hendecourt 1988; Allamandola, Tielens \&
Barker 1989), and the source of these aromatic emission features is widely identified with
polycyclic aromatic hydrocarbon (PAH) molecules, though other assignments have been made
\citep{sw89}.  The mid--infrared emission features have been observed in a wide
variety of astrophysical environments including external galaxies and so we include a PAH
component in our model.

With the grain composition established for our modeling purposes, we need to specify the
absorption, scattering, and extinction cross sections, $\sigma_{abs,sca,ext}$ of the
grains as well as the their size distribution and abundances.  For spherical grains,
$\sigma(a,\lambda)$ is specified in terms of the efficiencies, $Q(a,\lambda)$;

\begin{equation}
\label{eq:cs_def}
\sigma(a,\lambda) = \pi a^{2} Q(a,\lambda), 
\end{equation}

\noindent
where $a$ is the radius of the grain.  The optical scattering and absorption efficiencies
of the graphite, amorphous carbon (AMC), and silicate grain populations were derived from
Mie theory \citep{bh83} with the dielectric functions described in \citet{ld93}
(graphite), \citet{zmcb96} (AMC), and \citet{wd00} (silicate). The cross--sections for the
PAH molecules were derived from the analytic form presented in \citet{d97}.  This analytic
form is in turn based on the work of \citet{lhd89} and D\'{e}sert, Boulanger \& Puget
(1990) who decompose the PAH cross--section into three parts; UV-visual continuum, IR
continuum, and IR lines.  The PAH cross--section is based both on laboratory data
(UV--visual; IR line integrated cross sections) and observations of the reflection nebula
NGC~2023 (IR line widths and continuum).  These PAH cross--sections do not include a
2175~\AA\ bump, whereas many laboratory data suggest that PAH molecules do have absorption
features in the UV, though they have difficulty reproducing the observed stability of the
central wavelength of the bump.  In light of the uncertainty in constructing dust grain
models, we do not consider this a serious problem; as greater understanding of dust grain
behavior becomes available (including, for example, accounting for the non-bulk nature of
the optical properties of nano-sized grains and molecules), our input dust model can be
easily extended to include these sorts of refinements. Since the origin of some of the PAH
emission line features is in C--C modes and others in C--H modes, their relative strengths
can be adjusted by allowing the hydrogen coverage $(x_H \equiv H_{present}/H_{sites})$ to
vary \citep{pl89}.  As our purpose here is not the detailed fitting of individual objects
nor the prediction of the strengths of individual aromatic features, we set $x_H = 1$ and
do not investigate the effects of varying it further.

Mathis, Rumpl \& Nordseick (1977, MRN) showed that the near IR to far UV extinction curve
could be reproduced by a simple two component (graphite + silicates) dust model with a
power law size distribution, $n(a) \propto a^{-3.5}$.  The original MRN model included
grain sizes from $a_{min} = 50$~\AA\ to $a_{max} = 0.25~\micron$.  The large lower limit on
the size of the grains in the MRN model is a serious limitation.  The grains are large
enough that they almost always maintain equilibrium temperatures which are too low to
explain the observed emission shortward of $\sim 60~\micron$ in the Milky Way and other
galaxies.  Modeling the near to mid-IR emission requires the inclusion of a population of
small grains that undergo temperature fluctuations and hence spend some fraction of the
time emitting at temperatures in substantial excess of their equilibrium temperature (see
\S \ref{sec:trans}).  In our model, the graphite, AMC, and silicate grains have radii that
extend from $a_{min} = 8.5$\AA\ to $a_{max} = 3 \micron$. 

Since the grain optical constants used to derive the scattering, emission, and absorption
efficiencies of the grains do not vary depending on the extinction curve, reproducing the
observed extinction curve features in the different environments requires that we vary the
size distributions and relative abundances of the different grain species.  We derive the
size distributions, $dn(a)/da$ (grains $\micron^{-1}$ H$^{-1}$), of the graphite, AMC, and
silicate grains from fits to the observed extinction curves in various environments,
including the Milky Way, and the Large and Small Magellanic Clouds (MW, LMC, \& SMC). We
have constructed two dust models for each extinction curve, one consisting of a population
of graphite and silicate grains (A) and the other of graphite, AMC, and silicate grains
(B).  The observed extinction curves are fit using the maximum entropy method which seeks
to find the smoothest size distributions consistent with the observations
\citep{kmh94,cwgm00,wolff+01}.  
The size distributions and relative abundances of the grain components are adjusted to
best fit the overall shape of the observed extinction curves.  For example, since our
grain model identifies the carriers of the 2175~\AA\ feature as small graphite grains, the
presence of the 2175~\AA\ feature in the Milky Way extinction curve requires a large
population of these carriers.  The population of small graphite grains in the SMC will be
reduced with respect to the MW to reproduce the observed absence of the 2175~\AA\ 
extinction feature.  On the other hand, the paucity of small graphitic grains in our SMC
dust models requires a large population of small silicate grains in order to reproduce the
steep far UV rise in the SMC extinction curve.  Small silicate grains will be even more
important in the three component (silicate, AMC, and graphite) dust model for the SMC
since essentially all the carbon is amorphous in our dust model and does not contribute to
the far UV rise.  The observed extinction curves were taken from the literature: The MW
curve is taken to be the average MW curve as parameterized by Cardelli, Clayton \& Mathis
(1989) with $R_V = 3.1$.  We fit two average LMC extinction curves, one derived from
observations near the superbubble LMC~2 and the other from observations in the rest of the
LMC \citep{mcg99}.  The SMC extinction curve is taken from \citet{gc98}.

Models A and B are both extended to include a PAH component. The PAH component is included
as an extension of the carbonaceous component grain size distribution to a minimum size of
4~\AA\ ($\sim 20$ carbon atoms; $a_{PAH} \simeq 0.9 \sqrt{N_C}$~\AA).  At the upper end of
the PAH size distribution, we require that the number of carbon atoms in the largest PAH
molecule equal the number of carbon atoms in the smallest carbonaceous grain.  For
example, a spherical graphite grain of radius 8.5~\AA\ contains $N_C = (a/1.29)^3 \simeq
300$ carbon atoms, which we take as the number of carbon atoms in the largest PAH
molecule, corresponding to a maximum PAH size of $\sim 16$~\AA.  The PAH are tied to the
graphite grain size distribution and the AMC grain size distribution for models A and B,
respectively. Although the exact shape of the size distribution of small grains and
molecules is not well known and not well constrained by extinction curve fitting, a
substantial population is required, both to reproduced the observed extinction as well as
the mid IR emission from dust; for the PAH size distribution, we assume a log-normal form
given by

\begin{equation}
\frac{dn(a)}{da} = \frac{A}{a} \mathrm{e} ^{-2.0 \left[ \ln \left(\frac{a}{4}\right)\right]^2}
\label{eq:pah_dnda}
\end{equation}

\noindent
(Weingartner \& Draine 2000) where $A$ is a normalization which we derive by
requiring that the graphite and PAH size distributions merge smoothly at $N_C = 300$,
i.e.,

\begin{equation}
\left. \left[
\left( \frac{dn}{dN_C} \right) _{PAH} = \left( \frac{dn}{dN_C} \right) _{Gr/AMC} \right]
\right| _{N_C = 300}.
\label{eq:pah_norm}
\end{equation}

\begin{figure}[t]
\plottwo{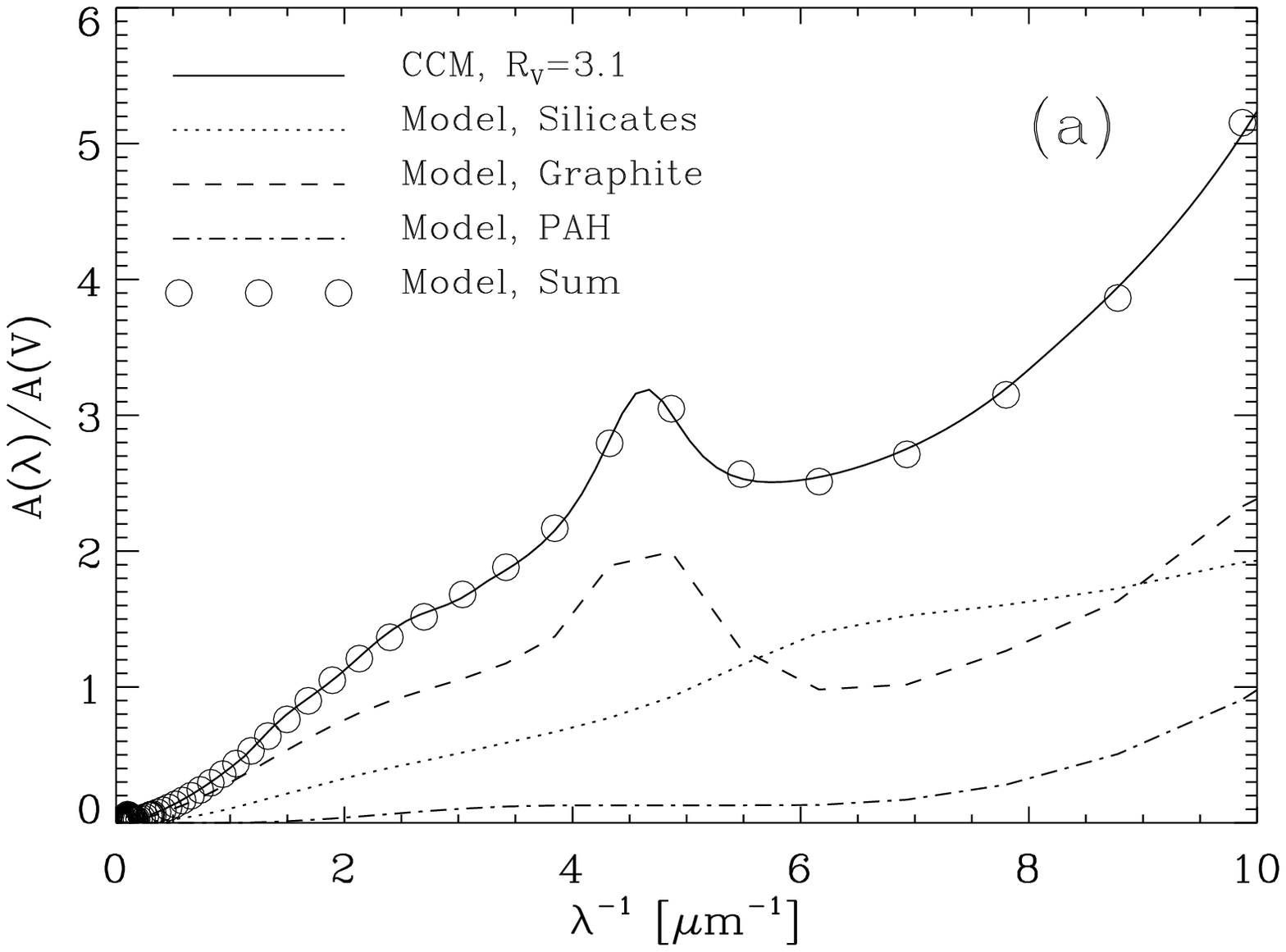}
        {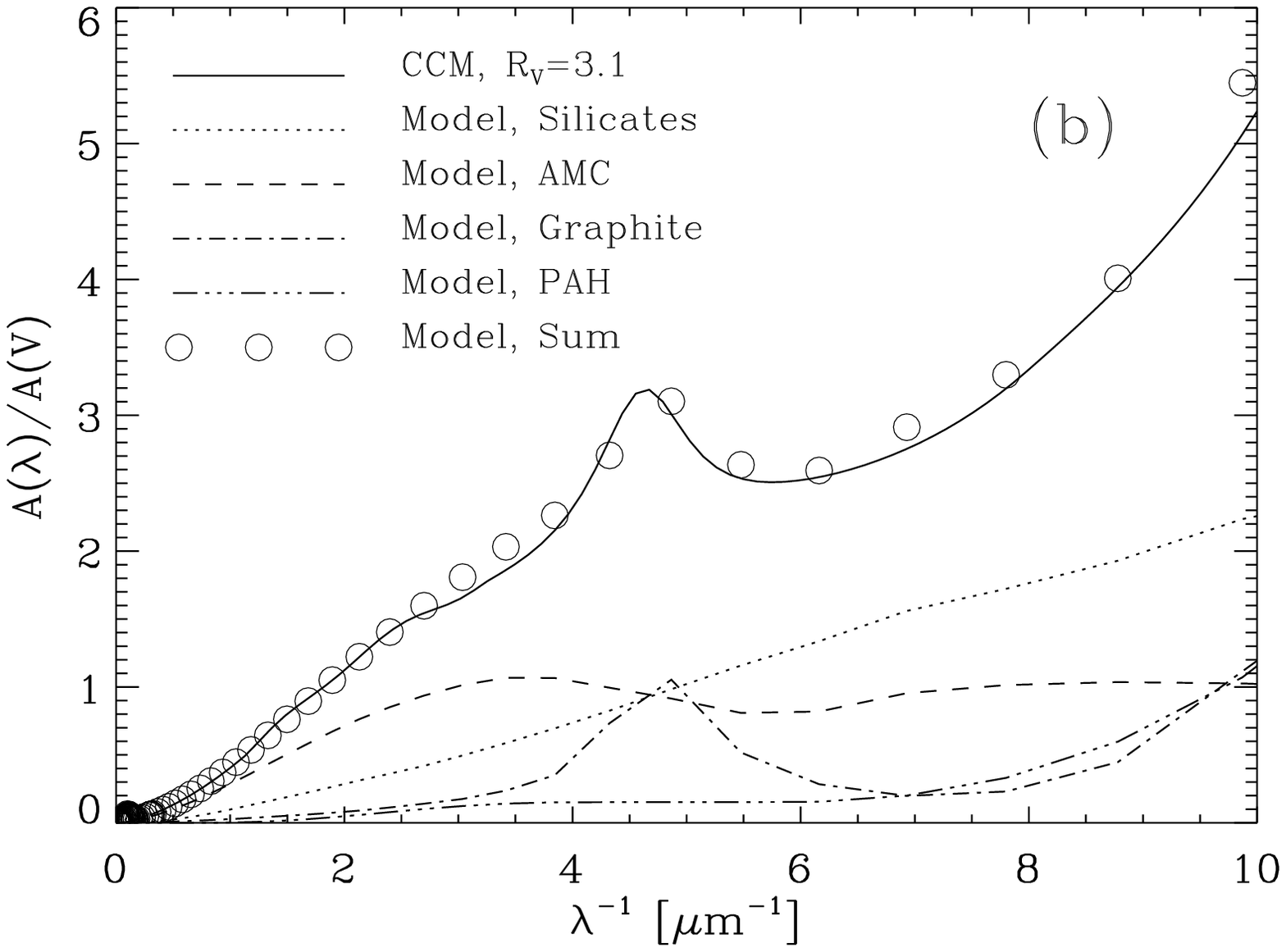}
\caption{Extinction curves for the adopted Milky Way dust models.
  (a) Three component (silicate, graphite, and PAH; Model A) (b) Four component (silicate,
  AMC, graphite, and PAH; Model B).  The contribution of each dust grain component to the
  extinction is plotted along with a CCM \citep{ccm} curve with R$_V$ =
  3.1. \label{fig:ext_mw}}
\end{figure}

\begin{figure}
\plottwo{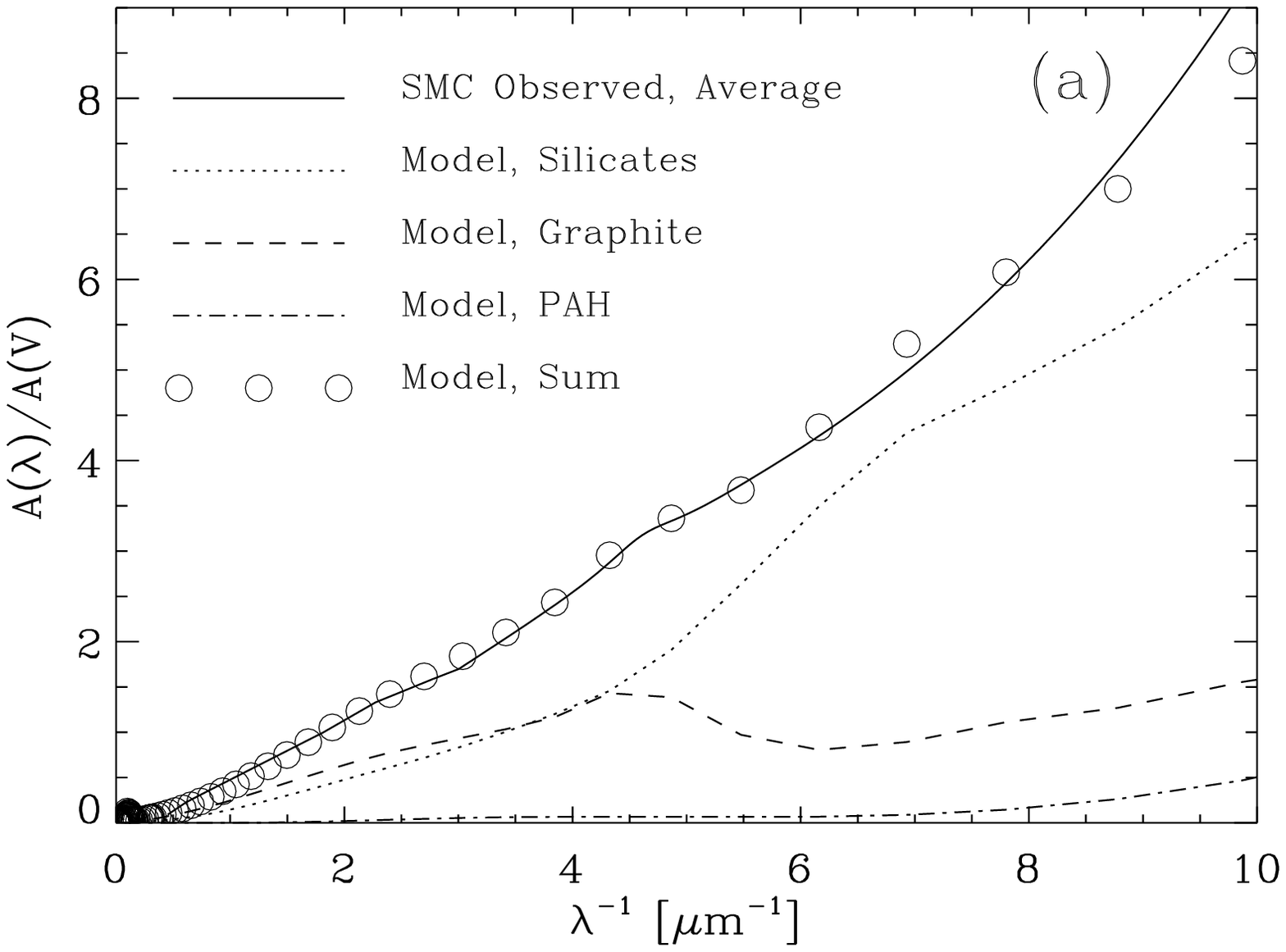}
        {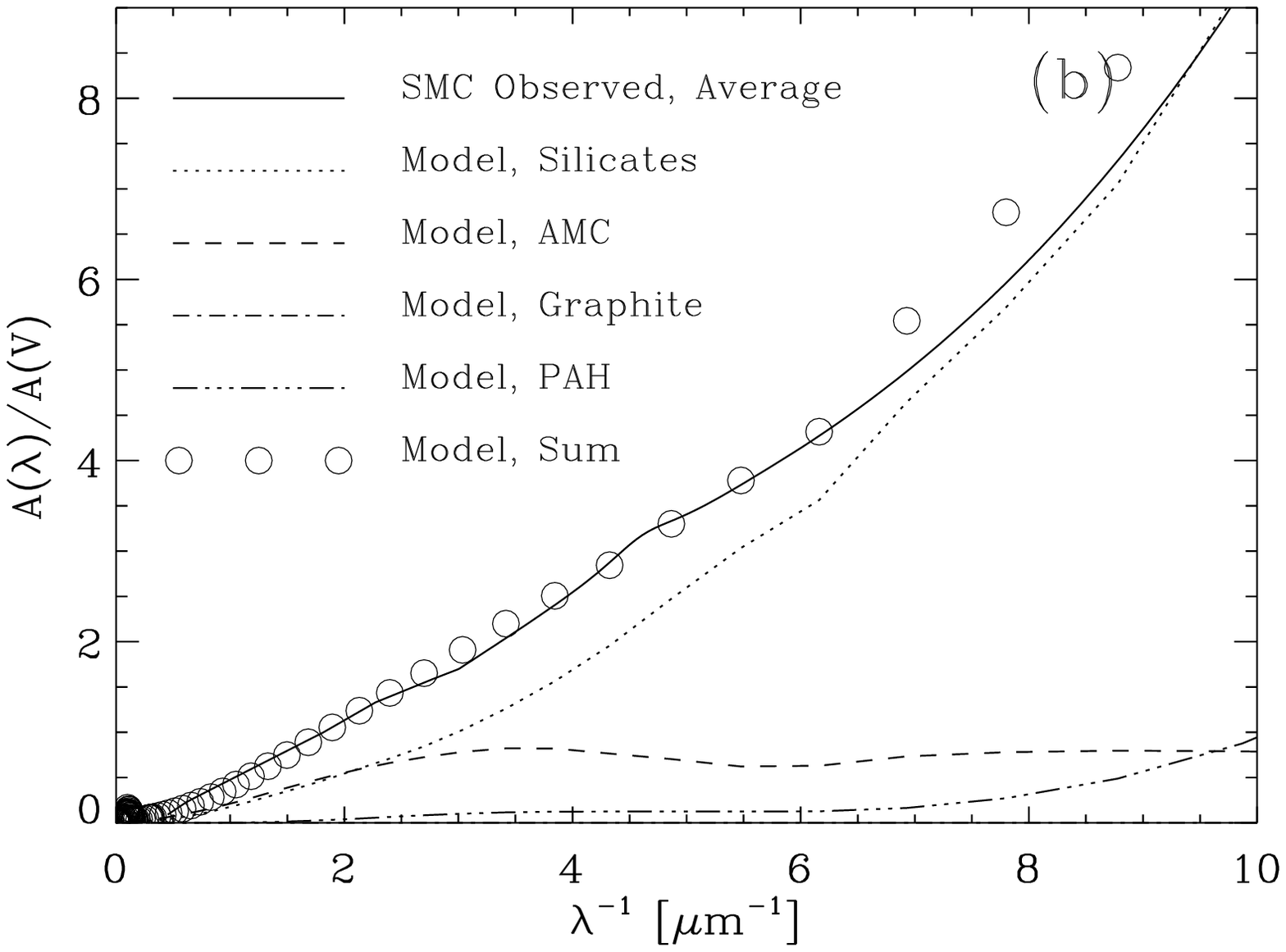}
\caption{Same as Figure \ref{fig:ext_mw} except for our adopted SMC dust models. The
  observed average SMC extinction curve is taken from \citet{gc98}. \label{fig:ext_smc}}
\end{figure}

\noindent
After including the PAH component, the size distributions of the carbonaceous grains are
normalized to insure conservation of the total mass of carbon in the dust model.  In Table
\ref{tbl:dust_abundance}, we report the abundances of each grain component in the
four dust models we consider (MW, LMC, LMC~2, \& SMC).  In Figures \ref{fig:ext_mw} and
\ref{fig:ext_smc}, we show the observed extinction curves for the MW and SMC,
respectively, along with our model predictions.  The model extinction curves have been
decomposed into the contributions from the three grain components.

\begin{figure}
\plotone{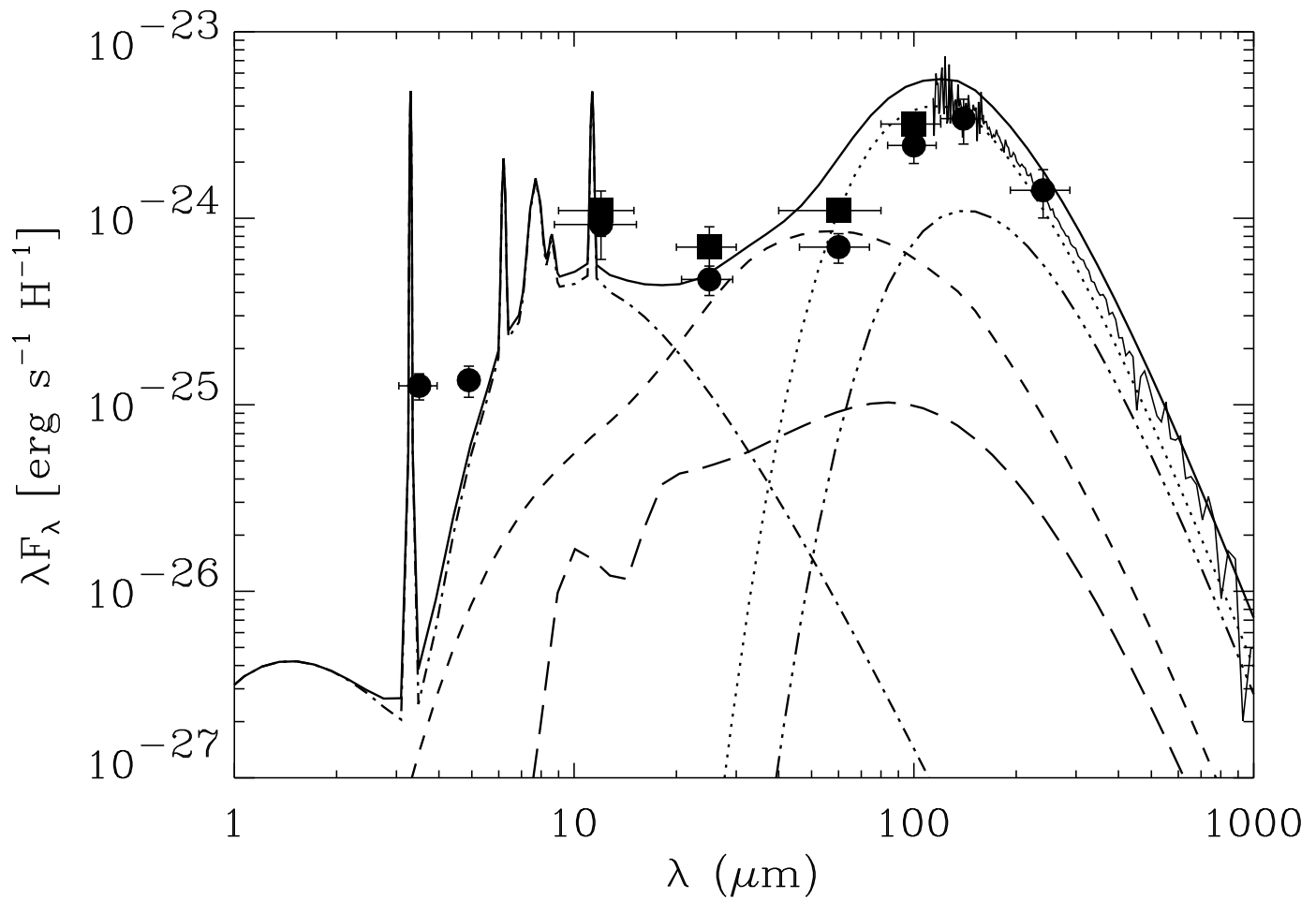}
\caption{Predicted emission spectrum from our MW dust model exposed to the local interstellar
  radiation field and compared to the observed diffuse ISM spectrum.  The radiation field is
  taken from \citet{mmp83}.  The predicted model spectrum (solid line) is decomposed into
  emission components from PAH molecules (dot-dash line) large ($>$100~\AA) graphite and
  silicate grains (dotted and triple dot-dash lines, respectively) and small ($\le$100~\AA) 
  graphite and silicate grains (short dash and long dash lines, respectively). The IRAS
  data (squares) are taken from \citet{bp88}; DIRBE (circles) and FIRAS (solid line) data
  are taken from \citet{d97}. \label{fig:dust_spectrum_isrf}}
\end{figure}

A grain model that is able to fit the observed extinction curves should also provide an
acceptable fit to the observed emission.  We have calculated the emission expected from
our dust model when exposed to the local radiation field as a check on the model.  The
local radiation field was taken from Mathis, Mezger, \& Panagia (1983). As can be seen in Figure
\ref{fig:dust_spectrum_isrf}, our MW grain model reproduces the diffuse interstellar IR
emission reasonably well from $\sim 3-1000~\micron$ though it overestimates by roughly 
a factor of two the emission in the 60, 100, and 140~\micron\ bands.  This disagreement 
is not unexpected as we have optimized our grain model to fit the observed average diffuse
extinction, not the emission; the emission and extinction are observed along different
lines of site and the dust grain populations are not necessarily the same.  

\begin{deluxetable}{llcccc}
\tablewidth{0pt}
\tablecaption{Model Dust Abundances \label{tbl:dust_abundance}}
\tablehead{
        &  & \colhead{MW} & \colhead{LMC} & \colhead{LMC~2} & \colhead{SMC}
} 
\startdata
        & Z$_{si}$\tablenotemark{a,b} & $5.8\times 10^{-3}$ & $1.9\times 10^{-3}$ &
                                        $2.0\times 10^{-3}$ & $1.1\times 10^{-3}$ \\

        & Z$_{gr}$\tablenotemark{a}   & $3.6\times 10^{-3}$ & $7.6\times 10^{-4}$ &
                                        $7.7\times 10^{-4}$ & $2.4\times 10^{-4}$ \\

Model A & Z$_{PAH}$\tablenotemark{a}  & $2.1\times 10^{-4}$ & $4.8\times 10^{-5}$ &
                                        $4.2\times 10^{-5}$ & $9.3\times 10^{-6}$ \\

        & [C/H]\tablenotemark{c} & 320 & 68 & 68 & 21 \\
        & [Si/H]\tablenotemark{c} & 34 & 11 & 12 & 6 \\
        & & & & & \\
\hline
        & & & & & \\
        & Z$_{si}$\tablenotemark{a,b} & $5.7\times 10^{-3}$ & $1.9\times 10^{-3}$ &
                                        $2.1\times 10^{-3}$ & $1.4\times 10^{-3}$ \\

        & Z$_{AMC}$\tablenotemark{a}  & $2.2\times 10^{-3}$ & $4.2\times 10^{-4}$ &
                                        $5.0\times 10^{-4}$ & $1.4\times 10^{-4}$ \\ 

Model B & Z$_{gr}$\tablenotemark{a}   & $8.1\times 10^{-4}$ & $2.6\times 10^{-4}$ &
                                        $1.8\times 10^{-4}$ &  \nodata  \\

        & Z$_{PAH}$\tablenotemark{a}  & $2.2\times 10^{-4}$ & $4.7\times 10^{-5}$ &
                                        $5.0\times 10^{-5}$ & $1.5\times 10^{-5}$ \\

        & [C/H]\tablenotemark{c} & 271 & 61 & 61 & 13 \\
        & [Si/H]\tablenotemark{c} & 33 & 11 & 12 & 8 \\

\enddata

\tablenotetext{a}{Fraction by mass relative to hydrogen.}
\tablenotetext{b}{Assuming a mixture of crystalline fosterite
  (Mg$_2$SiO$_4$) and fayalite (Fe$_2$SiO$_4$) for the silicate dust.}
\tablenotetext{c}{Abundance of carbon and silicon (assuming
  Mg$_2$SiO$_4$ and Fe$_2$SiO$_4$ in equal parts) relative to hydrogen, in PPM.}
\end{deluxetable} 

\section{Dust Heating and Emission}
\label{sec:physics}
In this section we describe the derivation of the dust emission spectrum given a dust 
grain model and heating sources.  The determination of the emission spectrum reduces to 
the problem of determining the temperature dust grains of a given size and composition 
will reach when exposed to a radiation field.  We outline the relevant equations for 
both equilibrium and single--photon, or transient, heating of dust grains.  The actual
implementation of our heating code is discussed in \S\ref{sec:comp_method}.

\subsection{Equilibrium Heating}
\label{sec:equi}

The monochromatic energy absorbed by a spherical dust grain of radius $a$ and species $i$
exposed to a radiation field $J_{\lambda}$ is given by

\begin{equation}
E^{abs}_{i}(a,\lambda) = 4\pi \: \sigma_{i}(a,\lambda) \: J_{\lambda}
\label{eq:eq_ab_spec_ia}
\end{equation}

\noindent
where $\sigma_{i}(a,\lambda)$ is the absorption cross section. The energy emitted by the
same particle can be expressed by

\begin{equation}
E^{em}_{i}(a,\lambda) = 4\pi \: \sigma_{i}(a,\lambda) \: B_{\lambda}(T_{i,a})
\label{eq:eq_em_spec_ia}
\end{equation}

\noindent
where $B_{\lambda}(T_{i,a})$ is the Planck function evaluated at the temperature of
the grain.  Thus, in a volume of space within which it is assumed $J_{\lambda}$ is
constant, the equation describing the equilibrium between the energy absorbed and emitted
by a single grain of radius $a$ can be written

\begin{equation}
\int\limits_{0}^{\infty} \! d\lambda \;\; \sigma_{i}(a,\lambda) \: J_{\lambda}  = 
\int\limits_{0}^{\infty} \! d\lambda \;\; \sigma_{i}(a,\lambda) \: B_{\lambda}(T_{i,a}).
\label{eq:eq_energy_balance_ia}
\end{equation}

\noindent
Eq. (\ref{eq:eq_energy_balance_ia}) can be solved iteratively for the equilibrium
temperature of the dust grain.  With the temperature of each individual dust grains known,
the dust emission spectrum from all species and grain sizes is calculated in a
straightforward manner:

\begin{equation}
L(\lambda) = 4\pi \sum_{i} \int\limits_{a_{min}}^{a_{max}}\! da \;\; n_{i}(a) \:
\sigma_{i}(a,\lambda)  \: B_{\lambda}(T_{i,a}).
\label{eq:eq_dust_spectrum}
\end{equation}

\subsection{Transient Heating}
\label{sec:trans}
The main assumption in the above discussion is that the dust grains reach equilibrium with
the radiation field and can be characterized by a single temperature.  While this
assumption is valid for large grains, it is not generally true for small grains.  The
absorption of a single high energy photon by a small grain can raise its temperature
significantly above its equilibrium temperature. Which grains fall into the ``large'' and
``small'' categories depends on the grain composition as well as the characteristics of
the radiation field, for computational purposes the division can be roughly made at $a
\simeq 100$~\AA.  For example, a 100~\AA\ graphite grain at a temperature of 25~K which
absorbs a Lyman limit photon ($\lambda = 912$~\AA) is heated to $\sim 39$~K, while a
40~\AA\ grain will reach a temperature of $\sim 90$~K, with the temperature change
becoming progressively larger for smaller and smaller grains.  This temperature represents
the maximum temperature the grain can reach. So while grains of all sizes will
have time dependent temperatures characterized by a probability distribution, $P(T)$,
rather than a single temperature, $P(T)$ will be narrowly distributed about the
equilibrium temperature for large grains but small grains will have very broad temperature
probability distributions.  In this case, the Planck function in Eq.
\ref{eq:eq_dust_spectrum} must be replaced by an integral over the probability
distribution, $P(T)$;

\begin{equation}
L(\lambda) = 4\pi \sum_{i} \int\limits_{a_{min}}^{a_{max}}\! da \;\; n_{i}(a) \:
\sigma_{i}(a,\lambda)  \: \int \! dT \;\; B_{\lambda}(T_{i,a}) \: P(T_{i,a}).
\label{eq:tr_dust_spectrum}
\end{equation}

\noindent
It can be seen that Eq. \ref{eq:eq_dust_spectrum} is a special case of Eq.
\ref{eq:tr_dust_spectrum} with $P(T) = \delta(T-T_{eq})$.  In determining $P(T)$, we follow
the method of \citet{gd89}.  They define a transition matrix $A_{f,i}$
whose elements are the probabilities that a grain undergoes transitions between
arbitrarily chosen internal energy states $i$ and $f$.  Determining $P(T)$ then amounts to
solving the matrix equation

\begin{equation}
\sum_{i=1}^N A_{f,i}P_i = 0.
\label{eq:matrix_eq}
\end{equation}

\noindent
The elements of $A_{f,i}$ are given by, in the case of heating ($f > i$)

\begin{equation}
A_{f,i} = 4\pi \: \sigma(a,\lambda) \: J_{\lambda} \frac{hc\Delta H_f}{(H_f - H_i)^3}
\label{eq:matrix_heating}
\end{equation}

\noindent
where $H_f$ and $H_i$ are the enthalpies of the final and initial states respectively and
$\Delta H_f$ is the width of the final state.  In the case of cooling ($f < i$), the
elements of $A_{f,i}$ are given by

\begin{equation}
A_{f,i} = \left\{
\begin{array}{rcr}
  \frac{4\pi}{\Delta H_i} \int\limits_{0}^{\infty}\!d\lambda \;\; \sigma(a,\lambda) \:
  B_\lambda(T_i) & \qquad & i = f+1 \\
  0              & \qquad & \mbox{otherwise}
\end{array}
\right.
\label{eq:matrix_cooling}
\end{equation}

\noindent
The requirement $i=f+1$ unrealistically allows cooling transitions to occur only to the
next lower level. While this has little effect at short wavelengths ($<$40~\micron), it
can underestimate the emission from small grains at submm wavelengths (e.g., Siebenmorgen,
Krugel \& Mathis 1992). However, the solution of Eq. \ref{eq:matrix_eq} without the
assumption of Eq.  \ref{eq:matrix_cooling} would require the inversion of a large matrix
which is prohibitively expensive in computation time when incorporated in our radiative
transfer code.  As the long wavelength emission from dust is dominated by large grains in
our application and speed is crucial, we adopt Eq.  \ref{eq:matrix_cooling} and the
attendant fast solution to Eq.  \ref{eq:matrix_eq} outlined by \citet{gd89}. Note that in
the above discussion we consider only radiative processes, which are the most important
for the modeling considered here.  However, other sources of grain heating (e.g.,
electron-grain collisions) and cooling (e.g., photoelectric emission) can be modeled by
additional terms in Eqs. \ref{eq:eq_energy_balance_ia}, \ref{eq:matrix_heating} \&
\ref{eq:matrix_cooling} \citep{gd89}.

\begin{figure}
\plottwo{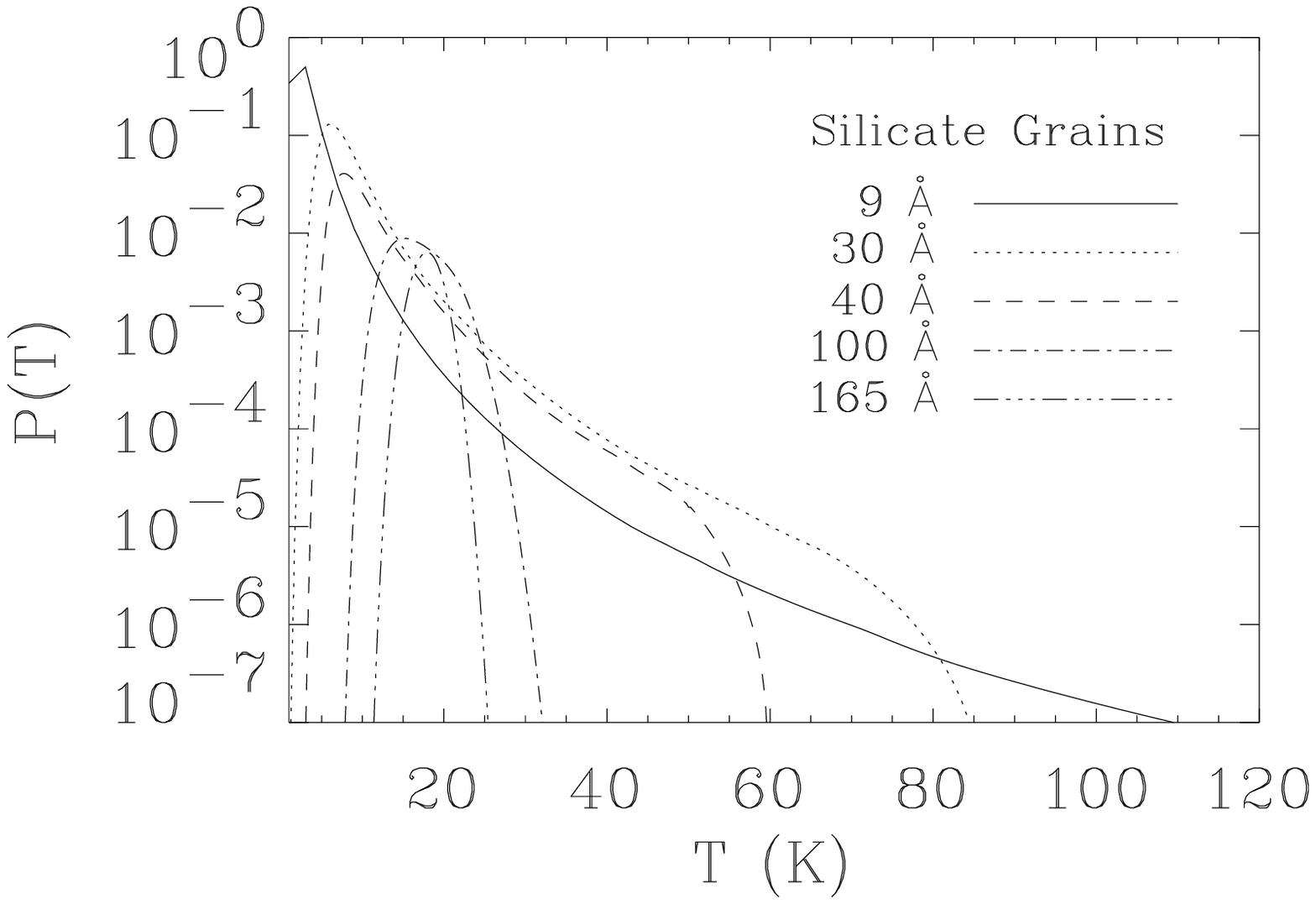}
        {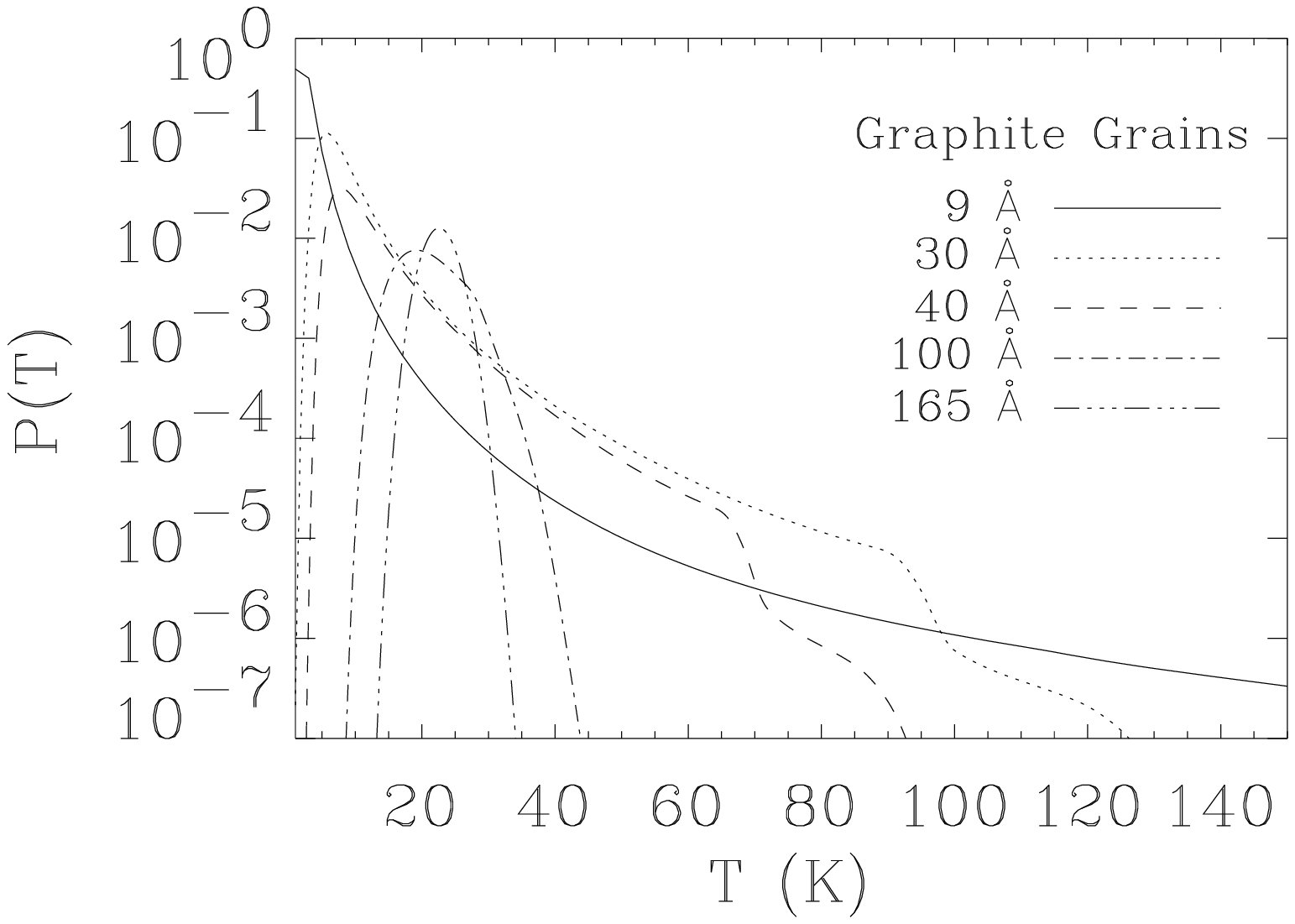}
\caption{Model calculations of the temperature probability distributions of silicate and
  graphite grains of various sizes when exposed to the local interstellar radiation field 
  \citep{mmp83}.
  \label{fig:prob_dist_isrf}}
\end{figure}

In Figure \ref{fig:prob_dist_isrf}, we show the grain temperature probability
distributions (Eq. \ref{eq:matrix_eq}) obtained from our algorithm when the grains are
placed in the Milky Way local radiation field (\S \ref{sec:dust_model}).  Although the
specifics of the probability distributions will vary with the radiation field, the general
behaviors exhibited in Figure \ref{fig:prob_dist_isrf} are characteristic (e.g.,
Siebenmorgen et al. 1992).  The smaller grains have very broad temperature distributions
indicating that they have a significant probability of reaching temperatures well above
and below their equilibrium temperatures.  As the grain size increases, the probability
distributions in a given radiation field narrow, approaching a delta function centered on
their equilibrium temperatures.  The width of the probability distribution is an
indication of whether the grain heating can be treated as an equilibrium process
(\S\ref{sec:comp_method}); the narrower the distribution the less important transient
heating effects become.

The PAH component of our dust model (\S \ref{sec:dust_model}) will also undergo
temperature fluctuations. If the cooling behavior of the molecule is known, the emission
from the molecules can be calculated using time averages \citep{lp84}.  The advantage of
this treatment over that of \citet{gd89} outlined above is an order of magnitude increase
in the speed of the calculation (Siebenmorgen, Krugel \& Mathis 1992). We treat the
transient heating of the PAH component via time averages assuming a time dependence for
the temperature of the molecule

\begin{equation}
T(t) = (T_p^{-0.4} + 0.005 t)^{-2.5}
\label{eq:pah_cooling}
\end{equation}

\noindent
where $T_p$ is the peak temperature reached by the molecule \citep{skm92}.  The mean time
between photon absorptions is calculated from 

\begin{equation}
\frac{1}{\overline{t}} = \frac{4\pi}{hc}\int\limits_0^{\lambda_c} \! d\lambda \: \lambda
\: \sigma (a,\lambda) \: J_{\lambda}
\label{eq:mean_time}
\end{equation}

\noindent
where $\lambda_c$ is the cut-off wavelength in the optical/UV cross-section of the PAH
molecule, defined by $\lambda_c = a/12.5$\micron, when $a$ is in \AA\ \citep{dbp90}.  The
existence of a cut-off wavelength in the optical/UV cross-section of the PAH results from
the discrete nature of the electronic levels in neutral PAH molecules \citep{dbp90}.  It is
assumed that the molecule absorbs a single photon of wavelength

\begin{equation}
\lambda_{abs} = \frac{\int\limits_0^{\lambda_c} \! d\lambda \;\; \lambda \:
\sigma (a,\lambda) \: J_{\lambda}}{\int\limits_0^{\lambda_c} \! d\lambda \;\;
\sigma (a,\lambda) \: J_{\lambda}}
\label{eq:mean_wave}
\end{equation}

\noindent
and cools following Eq. \ref{eq:pah_cooling} with $T_p$ calculated from the enthalpy of
the PAH molecule.  The PAH emission spectrum is then calculated
from 

\begin{equation}
L_{PAH}(a,\lambda) = 4\pi \: \sigma (a,\lambda) \: \overline{B[T(t)]} 
\label{eq:pah_spec}
\end{equation}

\noindent
where the bar indicates an average taken over the mean time between photon absorptions
(Eq. \ref{eq:mean_time}).  With the cooling behavior of the molecule approximated by Eq.
\ref{eq:pah_cooling}, energy conservation is not necessarily strictly maintained and we
re-normalize the emitted spectrum to ensure energy conservation \citep{skm92}.  Treating
the PAH emission in this manner is significantly faster than the matrix method of
\citet{gd89} outlined above for the small carbon and silicate grains \citep{skm92}.

As stated above, treating the transient heating of a grain or molecule, requires knowledge of their enthalpy 
(e.g., Eqs. \ref{eq:matrix_heating},\ref{eq:matrix_cooling},\ref{eq:pah_cooling}).  The enthalpy
of the grains at temperature $T$ is defined in terms of their specific heats, $C(T)$, through

\begin{equation}
H(T) = \int\limits_{0}^{T} \! dT \; C(T).
\label{eq:enthalpy}
\end{equation}

\noindent
Specific heats and enthalpies for the graphite and silicate grains were taken from
\citet{gd89}. The enthalpy for the AMC component was assumed to be the same as that of the
graphite grains.  The specific heat for the PAH molecules was taken from the linear
approximation of \citet{sgbd98} to the data of \citet{lhd89}.    

\subsection{Computational Method}
\label{sec:comp_method}
In order to compute the radiative transfer and dust emission for a system, we define the
spatial distributions of the gas, dust, and heating sources within an arbitrary three
dimensional model space.  To illustrate the dependence of the IR spectrum on various
parameters, here we consider a spatial grid in the shape of a cube divided into $N^{3}$
bins.  The number of model bins essentially establishes the smallest spatial scale of
inhomogeneity resolved by the model, since the ratio of clump size to system radius is
defined by $1/N$ \citep{wg96}.  For the models consider here, we adopt $N = 30$.  We
consider a two-phase clumpy medium consisting of high and low density clumps, where the
density of each bin is assigned randomly.  The frequency of occurrence of high density
clumps is determined by the filling factor ($f\!\!f$) and the relative density of high and
low density model bins is characterized by the density ratio, $k = k_2/k_1$ where $k_2$
and $k_1$ are the densities of the low and high density media, respectively.  In the
following, we describe our computational method.

A single run of our model (where by single run we refer to a single set of input
parameters, e.g., dust grain model, heating sources and their relative distribution, size,
optical depth, filling factor, and density ratio) consists of the following iterative
procedure:

\begin{enumerate}
  
\item Monte Carlo radiative transfer of the photons from stellar and nebular sources
  through the model space, resulting in the directly transmitted, scattered, and absorbed
  fractions of the initial input photons in each model bin. 

\item Calculation of the dust emission spectrum based on the heating supplied by the
  fraction of the input energy absorbed in the dust and the choice of dust model.
  
\item Monte Carlo radiative transfer of the emitted {\em dust} spectrum through the model
  space, resulting in a new grid of transmitted, scattered, and absorbed fractions.
      
\item Convergence check. If the fractional change in the absorbed energy grid from the
  previous iteration is less than some tolerance (we have taken $\delta = 0.01$),
  convergence is achieved and the run terminates.  Otherwise, return to step 2 with the
  new absorbed energy grid.

\end{enumerate}

\noindent
In the following, we will describe in some detail steps 2 and 4.  For a detailed
discussion of the Monte Carlo radiative transfer algorithm, the reader is referred to
\citet{gmwc00}.

From the Monte Carlo radiative transfer, we calculate $E_{abs,i}(\lambda)$, the total
energy absorbed in each model bin.  This grid of absorbed energy is passed to the dust
heating algorithm.  With reference to Eqs. \ref{eq:eq_energy_balance_ia} and
\ref{eq:matrix_heating}, we see that we require the specific intensity of the radiation
field, $J_{\lambda,i}$ in the $i^{th}$ bin in order to calculate the dust temperatures and
hence the dust spectrum in that bin.  In the following, we drop the subscript $i$ and
implicitly assume that the calculations described are to be done in {\em each} of the
$N^{3}$ bins.  In order to calculate $J_{\lambda}$ from the absorbed energy, we make use
of Eq. \ref{eq:eq_ab_spec_ia}.  Multiplying by the size distribution, integrating over
grain size and summing over species, the left hand side of Eq.  \ref{eq:eq_ab_spec_ia}
becomes the total energy absorbed by all grain components of all sizes and we can write

\begin{equation}
E_{abs}(\lambda) = 4\pi J_{\lambda} \overline{\sigma (\lambda)}
\label{eq:EJ}
\end{equation}

\noindent
where $\overline{\sigma (\lambda)}$ is the total cross section per
unit volume of the grain population: 

\begin{equation}
  \overline{\sigma (\lambda)} = \sum \int\limits_{a_{min}}^{a_{max}}\!\! da \; \frac{dn(a)}{da} 
  \sigma (a,\lambda)
  \label{eq:Total_CS}  
\end{equation}

\noindent
where the sum is over all grain components (i.e., graphite, silicates, and PAH molecules).
Thus, we derive the radiation field

\begin{equation}
J_{\lambda} = \frac{E_{abs}(\lambda)}{4\pi \overline{\sigma (\lambda)}}.
\label{eq:J}
\end{equation}

\noindent
With $J_{\lambda}$ known, we can proceed to deriving the temperature of each grain species and size 
in the equilibrium case or the temperature probability distribution in the transient case.   

The heating algorithm proceeds as follows.  The contribution of the PAH component to the
emitted spectrum in each bin is calculated from a straightforward application of Eqs.
\ref{eq:pah_cooling}--\ref{eq:pah_spec}.  For the graphite and silicate grains, the
calculation can be considerably more complicated.  For grains with sizes $a > 100$\AA, we
solve Eq.  \ref{eq:eq_energy_balance_ia} iteratively for the temperature and the
contribution of each grain size and species to the total dust spectrum in the bin is
calculated via Eq.  \ref{eq:eq_dust_spectrum}.  For grains with sizes $a \le 100$\AA, we
allow for the grain to undergo temperature fluctuations as described in \S\ref{sec:trans}.
Computationally, the grain temperature must be divided into a discrete mesh.  The
temperature is related to the enthalpy of the grains, so selecting the temperature mesh is
equivalent to defining the enthalpy mesh to be used in calculating the elements of the
transition matrix, $A_{f,i}$ (Eqs. \ref{eq:matrix_heating} \& \ref{eq:matrix_cooling}).
Care must be taken in defining the temperature mesh; the transition matrix in Eq.
\ref{eq:matrix_eq} is $N_T\times N_T$ where $N_T$ is the number of temperatures in the
mesh.  As solving Eq.  \ref{eq:matrix_eq} is the most computationally expensive part of
the code, it is advantageous to keep $N_T$ as small as possible.  However, the probability
distribution $P(T)$ must be well sampled, especially where it is changing rapidly.  Hence
it is crucial to select the temperature interval and $N_T$ carefully.  We adopt an
iterative approach to both the temperature interval selection as well as the number of
bins.  Considerable effort is made to set up the initial mesh carefully and our algorithm
incorporates as much {\em a priori} information about the behavior of $P(T)$ with grain
size as possible.  For example, referring to Figures \ref{fig:prob_dist_isrf}a,b, very
small grains have very broad, slowly varying temperature distributions and a broad, coarse
grid may be sufficient to determine $P(T)$.  On the other hand, as we near the transition
region between ``small'' and ``large'' grains, $P(T)$ becomes increasingly peaked near the
equilibrium temperature and the broad, coarse mesh is not sufficient to sample it well.
We start the algorithm by defining a relatively narrow, coarse mesh with $N_T = 50$
equally spaced temperature intervals centered on $T_{eq}$,

\begin{equation}
\begin{array}{rclcl}
  0.50~T_{eq} \le  & T & \le 1.50~T_{eq}  & \qquad & T_{eq} \le 100 \\
  T_{eq} - 100 \le & T & \le T_{eq} + 100 & \qquad & T_{eq} > 100.
\end{array}
\end{equation}

\noindent
Based on the behavior of $P(T)$ derived from Eq. \ref{eq:matrix_eq} with this temperature
grid, we adjust the upper and lower bounds of the temperature interval and the number of
enthalpy bins $N_T$. In the case of a large grain, the coarse initial mesh will be
insufficient to determine $P(T)$.  The failure of the coarse mesh is manifested in a
probability distribution that is highly peaked at low temperatures and zero elsewhere,
i.e. it approaches one in the lowest temperature interval and zero in all other
temperature intervals.  In this case, we increase the number of enthalpy bins, $N_T$, by
50\% and repeat the calculation to $P(T)$.  We repeat this procedure until $P(T)$ is well
behaved or we have exceeded the maximum number of enthalpy bins, $N_{T,max}$. $N_{T,max}$
is an input parameter that we have set to 800 for our model runs.  In practice,
$N_{T,max}$, is not exceeded in the initial set up; generally, 75 to 112 bins are
sufficient to sample $P(T)$ and begin testing for convergence for even the largest grains
treated by the transient heating algorithm (see below).  In the case of a small grain, $P(T)$
will be a smoothly varying function across the initial narrow temperature interval and
the interval needs to be expanded to insure we include the whole range of
temperatures that the grain has a non-negligible probability of achieving.  In this case, we
expand the temperature interval to extend from $T_{min}=2.7$~K to $T_{max}=2500$~K and,
keeping $N_T = 50$, recalculate $P(T)$.  This temperature interval brackets all likely
temperatures that the dust grain can reach. For very small grains, this coarse, broad
temperature mesh may be sufficient to begin testing for convergence.  However, for
intermediate sized grains, $P(T)$ may again become peaked at low temperatures and approach 
zero in the higher temperature bins. In this case, we define a new maximum temperature 

\begin{equation}
T_{max}^{new} = T_{max}^{old} + 0.5(T_{eq} - T_{max}^{old}).
\end{equation}

\noindent
$N_T$ is increased by 50\% and a new $P(T)$ is derived with the new temperature mesh.
This procedure is iterated until $P(T)$ is well behaved in the interval. In practice, one
to two iterations are sufficient to roughly establish the correct temperature interval for 
the grain. 

With the initial $P(T)$ determined as above, we calculate the predicted spectrum of the
transiently heated grain using Eq. \ref{eq:tr_dust_spectrum} and test for convergence.  We 
define the convergence of the transient heating algorithm based on a comparison of the
calculated total emitted energy and the absorbed energy,

\begin{equation}
\frac{\Delta E}{E_{abs}} = \frac{\left| E_{abs} - E_{em} \right|}{E_{abs}}
\label{eq:e_cons}
\end{equation}

\noindent
where $E_{abs}$ is calculated by integrating Eq. \ref{eq:eq_ab_spec_ia} over wavelength
and $E_{em}$ is calculated from 

\begin{equation}
E_{em} = 4\pi \int\limits_{0}^{\infty} d\lambda \;\; \sigma_{i}(a,\lambda) \: \int \! dT
\;\; B_{\lambda}(T_{i,a}) \: P(T_{i,a}).
\label{eq:tr_em_energy}
\end{equation}

\noindent
For convergence, we require that $\Delta E/E_{abs} < \delta E$, where $\delta E$ is an
input parameter that we have set to 0.1 for the runs presented here.  If the algorithm has not
converged, we define a new temperature interval, increase $N_T$ by 50\%, recalculate
$P(T)$ and re-evaluate $\Delta E/E_{abs}$.  Since there is no
advantage to including temperature intervals where the grain has a very small probability
of finding itself ($P(T) \ll 1$), we define a cutoff probability, $P_{cut} =
10^{-15}P_{max}$, where $P_{max}$ is the maximum of the probability distribution. The new
temperature interval is defined to exclude temperature bins for which $P(T) < P_{cut}$
\citep{mh98}.  This procedure is iterated until convergence is achieved or we exceed
$N_{T,max}$.  If $N_{T,max}$ is exceeded, the grain is treated as being at its equilibrium 
temperature and its contribution to the spectrum is calculated with
Eq. \ref{eq:eq_dust_spectrum}. The transient heating algorithm is turned off and all
subsequent grain sizes are treated via the equilibrium heating formalism.  

We have tested our algorithm for calculating the transient emission spectrum in a wide
variety of radiation fields from the local ISRF to the radiation field in close
proximity to a hot star to a variety of the SES models described above.  In all cases the
algorithm worked with no user interaction and produced probability distributions with the
correct behavior as a function of grain size and radiation field (e.g., see Figs.
\ref{fig:prob_dist_isrf}a,b). In addition, we have compared the results from our algorithm
with previous calculations in the literature \citep{skm92,mh98} with excellent
agreement.  In light of these tests, we are confident that we can apply our model to a
range of situations with minimal adjustments to the heating calculation.

A single run of the dust heating algorithm is complete when the above procedure has been
performed for all model bins for which the absorbed energy in that bin exceeds some cutoff
fraction.  The cutoff fraction is determined as follows.  In each bin we calculate the
fraction of the total energy absorbed,

\begin{equation}
f_{abs,i} = \frac{E_{abs,i}}{\sum_i E_{abs,i}}.  
\label{eq:frac_abs}
\end{equation}

\noindent
We then calculate the quantity $\sum _i f_{abs,i}$ for $f_{abs,i} > f_{cut}$ for a variety 
of values of $f_{cut}$.  We adopt a value of $f_{cut}$ such that the total energy absorbed 
in bins for which $f_{abs,i} > f_{cut}$ is larger than some target level of energy
conservation.  The target level is taken as an input and is generally between 0.95 and 1.
This level represents the best possible energy conservation that can be achieved for the
run; model bins with $f_{abs,i} < f_{cut}$ are not included in the dust heating
algorithm.  This procedure allows us to eliminate a large number of model bins in which
very little energy is absorbed, speeding up the calculation substantially with little cost in
accuracy.  

Upon completing a single run of the dust heating algorithm, we obtain the dust emission
spectrum from each point in the model.  In order to allow for the treatment of large
optical depths and to account for the dust self--absorption, we now add the dust as a new
source of emitted photons.  The Monte Carlo radiative transfer code is re--run, now using
the dust spectrum as the source of input photons rather than the stellar and nebular
sources.  The contribution of the dust emission to the absorbed energy in each model bin
as derived from the Monte Carlo is then added back into the absorbed energy grid from the
previous iteration, and the fractional change in the total absorbed energy is computed.
We consider the entire model run to have converged if the fractional change in absorbed
energy $\delta < 0.01$.  If $\delta \ge 0.01$, the new absorbed energy grid is passed to
the dust heating algorithm and steps 2--4 are repeated.  When convergence is achieved in
step 4, a single model run has been completed.  In general, even for large optical depths
(e.g., $\tau_V = 20 - 50$), convergence is achieved in fewer than 4 full iterations of
steps 2--4.  For optical depths in the range of $\tau_V = 2 - 10$, 2 iterations are
normally sufficient to reach convergence.

In addition to the far-UV to far-IR SED of the model, several other quantities are
included in the output upon the completion of a model run, including the size averaged
dust temperature for each dust component in each model bin, the relative fractions of the
total energy absorbed in the clump vs. interclump medium, and the total dust mass of the
model.  We define the size averaged dust temperature for each dust component in the
following way;

\begin{equation}
\label{eq:dust_temperature}
\overline{T}_i = \frac{\int\limits_{a_{min}}^{a_{max}}\! da \;\; n_{i}(a) T_{i}(a)}
                      {\int\limits_{a_{min}}^{a_{max}}\! da \;\; n_{i}(a)}.
\end{equation}

\noindent 
Note that in Eq. \ref{eq:dust_temperature}, $T_i$ is the equilibrium temperature of the
grain and hence the size averaged temperature does not account for the range of
temperatures reached by stochastically heated grains.  Since we have the temperature in
each bin, $\overline{T}$ may be used to calculate the radial temperature distribution of
the dust in the model (see \S \ref{sec:application}).

We calculate the total dust mass in each model in the following way:

\begin{equation}
\label{eq:dust_mass}
M_{dust} = \sum _{i}^{N_{bins}} \frac{N_{H,i}}{\tau _{V,i}} \left[ \tilde{\tau}_{V,i} V_i \sum_{j} 
\int\limits_{a_{min}}^{a_{max}}\!\! da \; \frac{4\pi}{3} a^3 \rho_{j} n_{j}(a) \right]
\end{equation}

\noindent
where the sum over $i$ is taken over model bins, $\tilde{\tau}_{V,i}$ is the optical depth
per unit $pc$ at $V$ in the $i^{th}$ bin and $V_i$ is the volume of the $i^{th}$ bin in
$pc^{3}$.  The term in brackets gives the dust mass per hydrogen column times
$\tau_{V,i}$, where the sum over $j$ is over dust components.  In order to calculate the
total dust mass we assume a gas to dust ratio and a value of the ratio of total to
selective extinction, $R_V$.  For each of the four dust models considered (see \S
\ref{sec:dust_model}), we take $R_V$ = 3.1 and a gas to dust ratio ($N_H/E(B-V))$of $5.8
\times 10^{21}$, $2.4 \times 10^{22}$, and $5.0 \times 10^{22}~~$ H
atoms/cm$^{2}$/mag$^{-1}$ for the MW, both LMC models, and the SMC, respectively.

\section{An Example: Starburst Galaxies}
\label{sec:application}

The \dm model is completely general in that we can treat arbitrary distributions of dust
and heating sources.  However, to illustrate how the model predictions depend on various
input parameters, here we apply \dm to input stellar distributions and geometric
environments appropriate for the modeling of starburst galaxies \citep{gcw97,wg00}.
\citet{wg00} defined different star/dust geometries including DUSTY and SHELL geometries.
The DUSTY geometry contains dust and stars mixed together, with both extending to the
system radius.  The SHELL geometry consists of stars extending to 0.3 of the system radius
with dust filling a shell extending from 0.3 to the system radius.  Pictorial
representations of these geometries can be found in Fig.  1 of \citet{wg00}.  We emphasize
that the DUSTY and SHELL geometries specify the global geometry of the model space;
locally, each model bin can be either clumpy or homogeneous, as characterized by the
$f\!\!f$ and the density ratio $k_2/k_1$.  While the model is capable of handling any
arbitrary geometry, these two geometries are expected to be representative of a wide range
of star/dust geometries, e.g., embedded stellar populations and mixed stars and dust.

Within the global geometries described above, MW, LMC, or SMC type dust (\S
\ref{sec:dust_model}) is distributed with a local geometry determined by the $f\!\!f$ and
density ratio.  The stellar population is distributed uniformly within the global
geometries.  The properties of the stellar population are taken from the stellar
evolutionary synthesis models of Fioc \& Rocca-Volmerange (1997, 2000;
http://www.iap.fr/users/fioc/PEGASE.html) (PEGASE models).  The starburst models presented
here were run using an IMF with a mass range of $0.1 - 100 M_{\sun}$ with a Salpeter
slope, a constant star formation scenario, ages ranging from $0-19$~Gyr, with a range of
metallicities from $-$2.3 to 0.7, and include a nebular component.  For a more detailed
discussion of the spectral synthesis model in the context of the \dm model, the reader is
referred to \citet{ghcrm99}.  The input stellar SEDs are rebinned to $\sim120$ wavelength
points.  The model SEDs for young input stellar populations exhibit features near 
0.37 and 0.67~\micron\ resulting from contributions to the rebinned SED from strong
nebular emission lines (O~II and H$\alpha$, N~II) in those wavelength regions.

In the following discussion, we present the results of several sets of \dm model runs
using the starburst stellar distributions and geometries described above.  We examine the
dependence of the spectrum, dust temperatures, and dust masses on variations in the input
parameters. The input parameters for \dm include: metallicity, star formation rate, and
age, which together determine the spectral shape and intensity of the input SED; the
filling factor ($f\!\!f$) and density ratio ($k=k_2/k_1$), which determine the clumpiness
of the scattering and absorbing dusty medium; the dust type which determines the
composition and size distribution of the dust grains; and the physical size of the region,
the global geometry, and the optical depth, $\tau_V$, which affect the dust mass as well
as the efficiency of a given mass of dust in absorbing the photons from the input SED.
The input optical depth, $\tau_V$, for both DUSTY and SHELL geometries is defined as the
radial optical depth, from the center to the edge of the model, that would result were the
dusty medium distributed homogeneously throughout the model space.  The range of values
for these parameters is tabulated in Table \ref{tbl:model_param}.  The case of the
application of \dm to starburst galaxies will be considered in more detail in a
forthcoming paper; in the following, we adopt dust Model A for illustrative purposes.

\begin{deluxetable}{lccc}
\tablewidth{0pt}
\tablecaption{Input Model Parameters \label{tbl:model_param}}
\tablehead{
  \colhead{Parameter} & \colhead{Range} & \colhead{Figure Reference} & \colhead{Fixed
    Value}\tablenotemark{a} 
}
\startdata
Metallicity & -0.4 & \nodata & \nodata \\
Dust Type & MW,SMC & \ref{fig:tr_onoff},\ref{fig:sed_mwsmc_comp}(a) & \nodata \\
Dust Model & A,B & \ref{fig:sed_mwsmc_comp}(b) & A \\
Global Geometry & SHELL,DUSTY & \ref{fig:sed_var_tv_cgg}, \ref{fig:temp_var_tv_cgg},
\ref{fig:sed_var_tv}, \ref{fig:dust_mass} & \nodata \\
$f\!\!f$ (filling factor) & 0.05 -- 0.5 & \ref{fig:var_k_ff} & 0.15 \\
$k$ (density ratio) & 0.001 -- 0.177 & \ref{fig:var_k_ff} & 0.01 \\
SFR & 0.5 -- 200~M$_{\odot}$~yr$^{-1}$ & \ref{fig:var_age_sfr},\ref{fig:cnsmass_comp} & 1.6~M$_{\odot}$~yr$^{-1}$ \\
Age & $10^{6}$ -- $19 \times 10^{9}$~yr & \ref{fig:var_age_sfr},\ref{fig:cnsmass_comp} & 40~Myr \\
Size & 10 -- 5000~pc & \ref{fig:var_size_sh} & 1000~pc \\
$\tau_V$ & 0.5 -- 20.0 & \ref{fig:sed_var_tv_cgg}, \ref{fig:temp_var_tv_cgg},
\ref{fig:sed_var_tv}, \ref{fig:dust_mass} & 10 \\
\enddata 

\tablenotetext{a}{Fixed value of parameter for model runs for which some other parameter
  is varied.}

\end{deluxetable}

\subsection{Transient Heating}
\label{sec:dis_trans}

\begin{figure}
\plotone{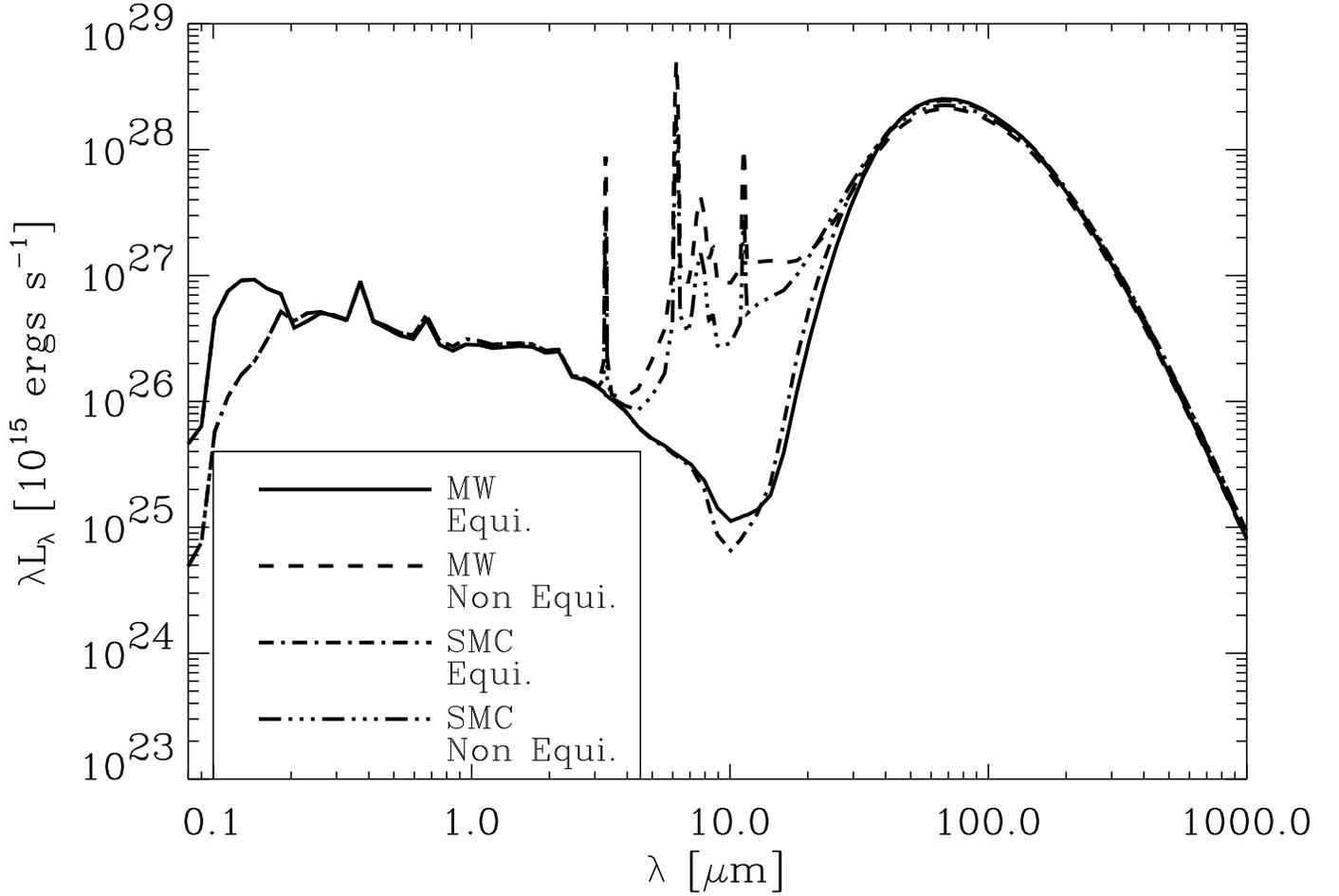}
\caption{Model SED's for MW and SMC type dust models with and without the effects of
  transient heating of the small grains.  The equilibrium cases include neither the
  effects of transient heating nor the emission from small grains nor a molecular (PAH)
  component, the non equilibrium cases include both.  All parameters are set to their
  values in column 4 of Table \ref{tbl:model_param} and a SHELL geometry is assumed.
  \label{fig:tr_onoff}}
\end{figure}

In Figure \ref{fig:tr_onoff}, we illustrate the effects of including the transient heating
(see \S\ref{sec:trans}) of small grains and molecules on the SEDs predicted by our
starburst model.  Including the effects of transient heating in the model increases the
predicted emission between $5-30$~\micron\ by as much as a factor of 20 for the cases
presented in Figure \ref{fig:tr_onoff}.

\subsection{Clumpiness}
\label{sec:dis_clumpy}
In this section, we illustrate the dependence of the predicted SED on the local structure
of the absorbing medium.  The local structure is characterized by $f\!\!f$ and $k$, and we
keep all other model inputs fixed at their values listed in column 4 of Table
\ref{tbl:model_param}, assuming a SHELL geometry and a model size of 30 bins per side.  We
have considered $f\!\!f$'s between $0.05$ and $0.5$, with a range of physical conditions
varying from an extended low density medium with rare, isolated high density clumps to an
interconnected network of high density clumps with a low density medium filling the voids.
The value of $k$ is varied between $0.001$ and $1$ (homogeneous) in steps of factors of
$\sim$5.62 ($k = 0.001,0.005,0.030,0.177,1.0$).  The effect of varying $f\!\!f$ and $k$ on
the SED is illustrated in Figs.  \ref{fig:var_k_ff}a,b,c,d.  We do not include the
homogeneous case in the Figures as the change in the SED between $k = 0.177$ and $1$ is
negligible for all values of $f\!\!f$.  For all values of $f\!\!f$, the effect of
increasing $k$ on the IR SED is to shift the peak of the dust emission to shorter
wavelengths, corresponding to higher dust temperatures.  The effect is quite small and
becomes less pronounced with increasing $f\!\!f$ and for $f\!\!f > 0.20$, the IR SED's are
essentially degenerate for different values of $k$.  The dominant reason for this behavior
is the increase in total energy absorbed by the dust with increasing $k$ at a constant
$f\!\!f$ which results in more heating and higher dust temperatures.  A smaller secondary
contribution to the behavior results from the fact that at higher values of $k$, a
correspondingly larger fraction of the energy is absorbed in the low density medium which
reaches higher temperatures than the dense clumps.  In any case, as can be seen from Fig.
\ref{fig:var_k_ff}, the model IR SED is not very sensitive to the local structure of the
medium.  The dominant effects of $f\!\!f$ and $k$ are seen in the optical and UV SED
\citep{wg96}.

\begin{figure}
\plottwo{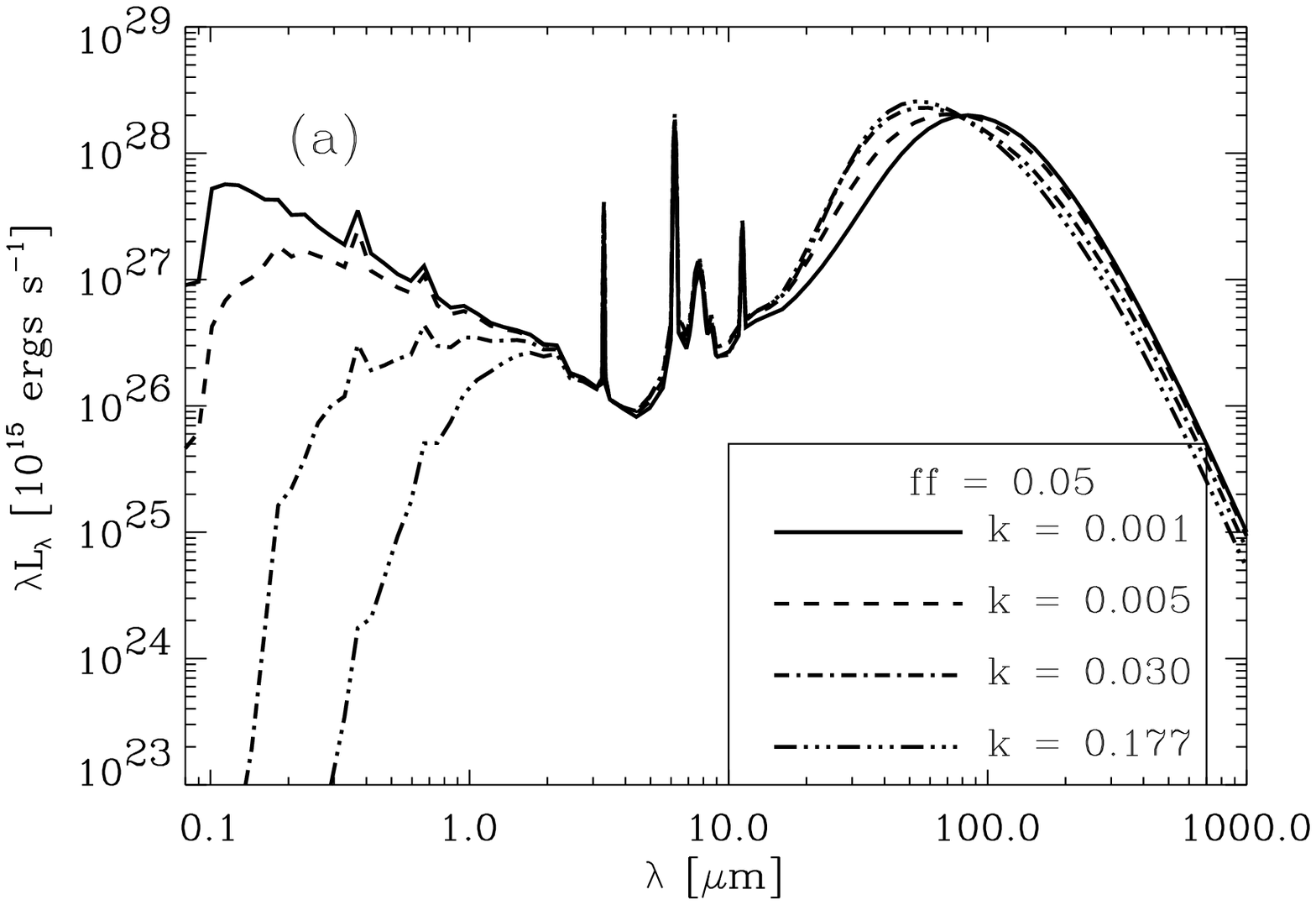}{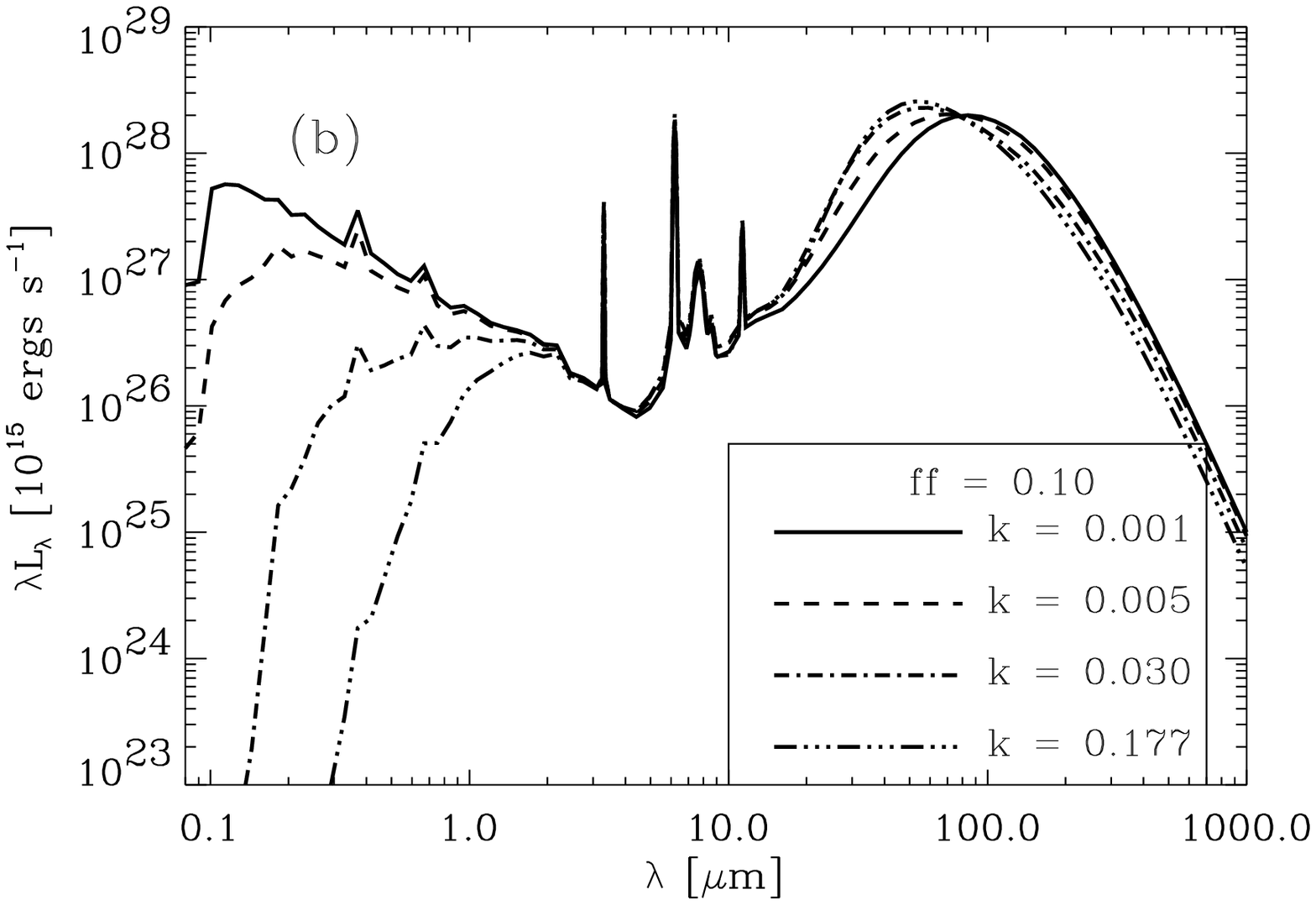}
\end{figure}

\begin{figure}
\plottwo{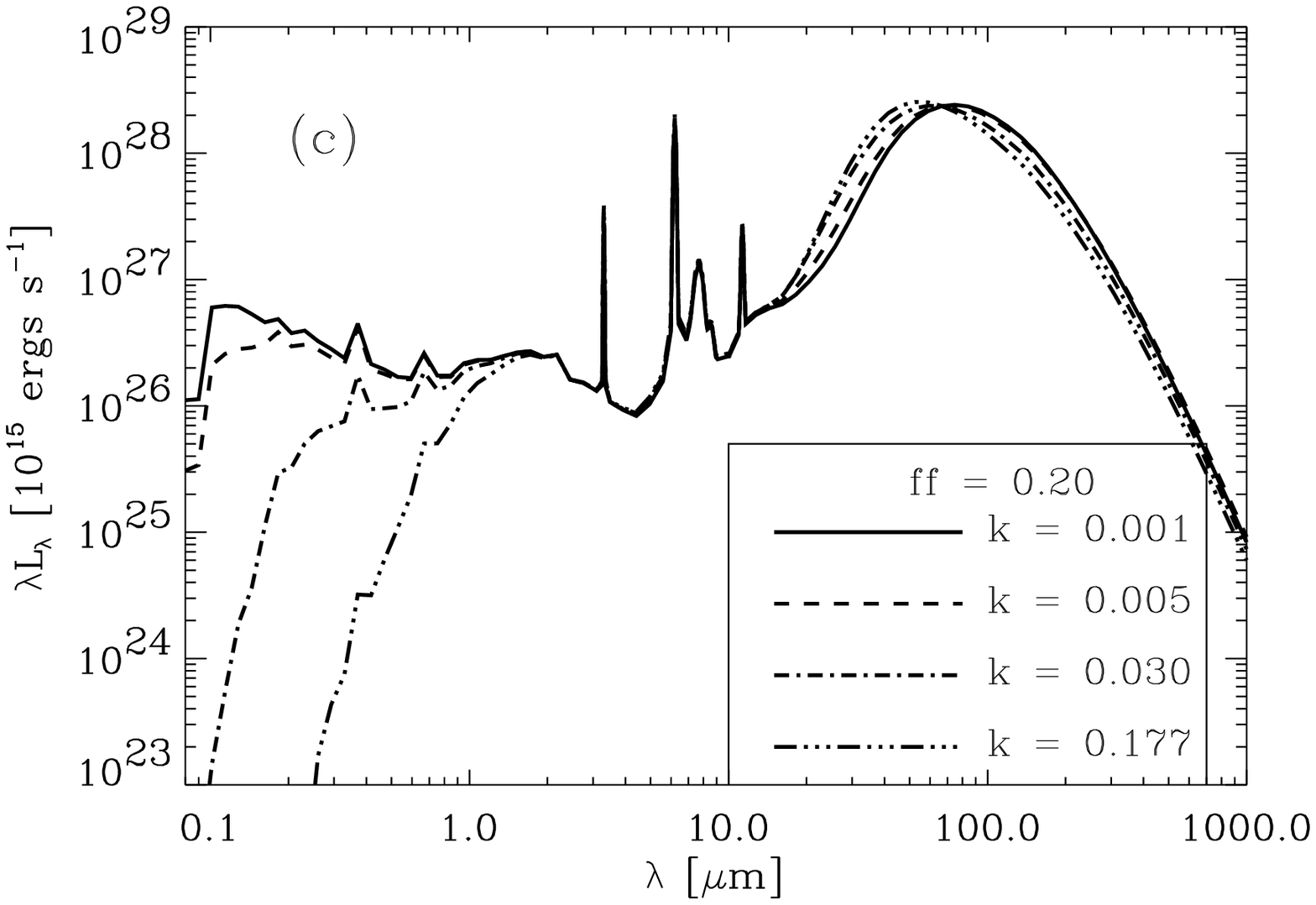}{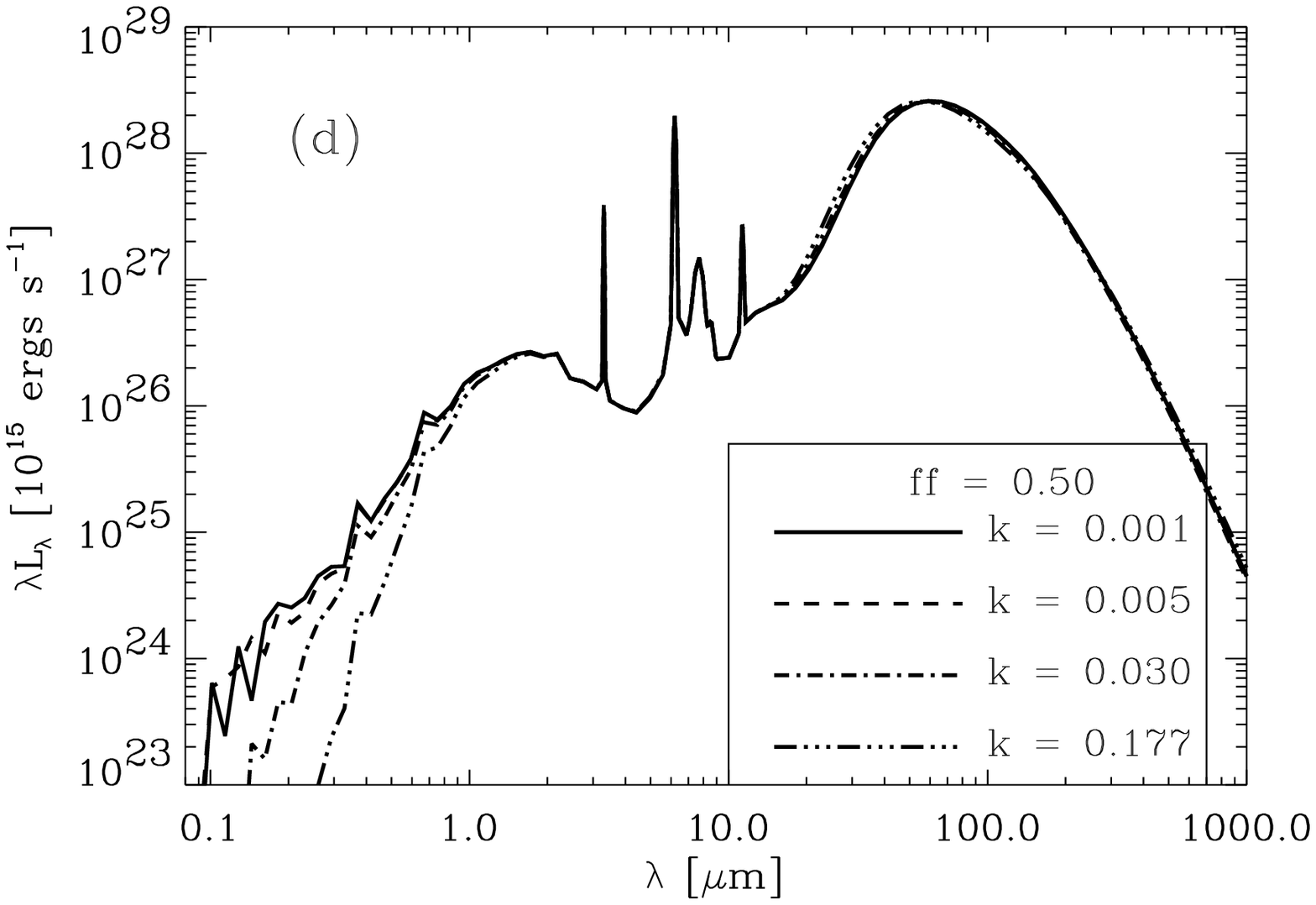}
\caption{Model SED's for a range of density ratios ($k = 0.001$ solid line, $k = 0.005$
  dashed line, $k = 0.03$ dash-dot line, $k=0.177$ dash-triple dot line) with different
  filling factors; (a) $f\!\!f=0.05$, (b) $f\!\!f=0.1$, (c)  $f\!\!f=0.2$, (d)
  $f\!\!f=0.5$. All models are calculated assuming SHELL geometry, 30 bins per side, and
  the remaining parameters specified in column 4 of Table  \ref{tbl:model_param}.
  \label{fig:var_k_ff}}
\end{figure}

\subsection{Dust Type (MW vs SMC) \& Dust Model (A vs B)} 
\label{sec:dis_dust}
\begin{figure}
\plottwo{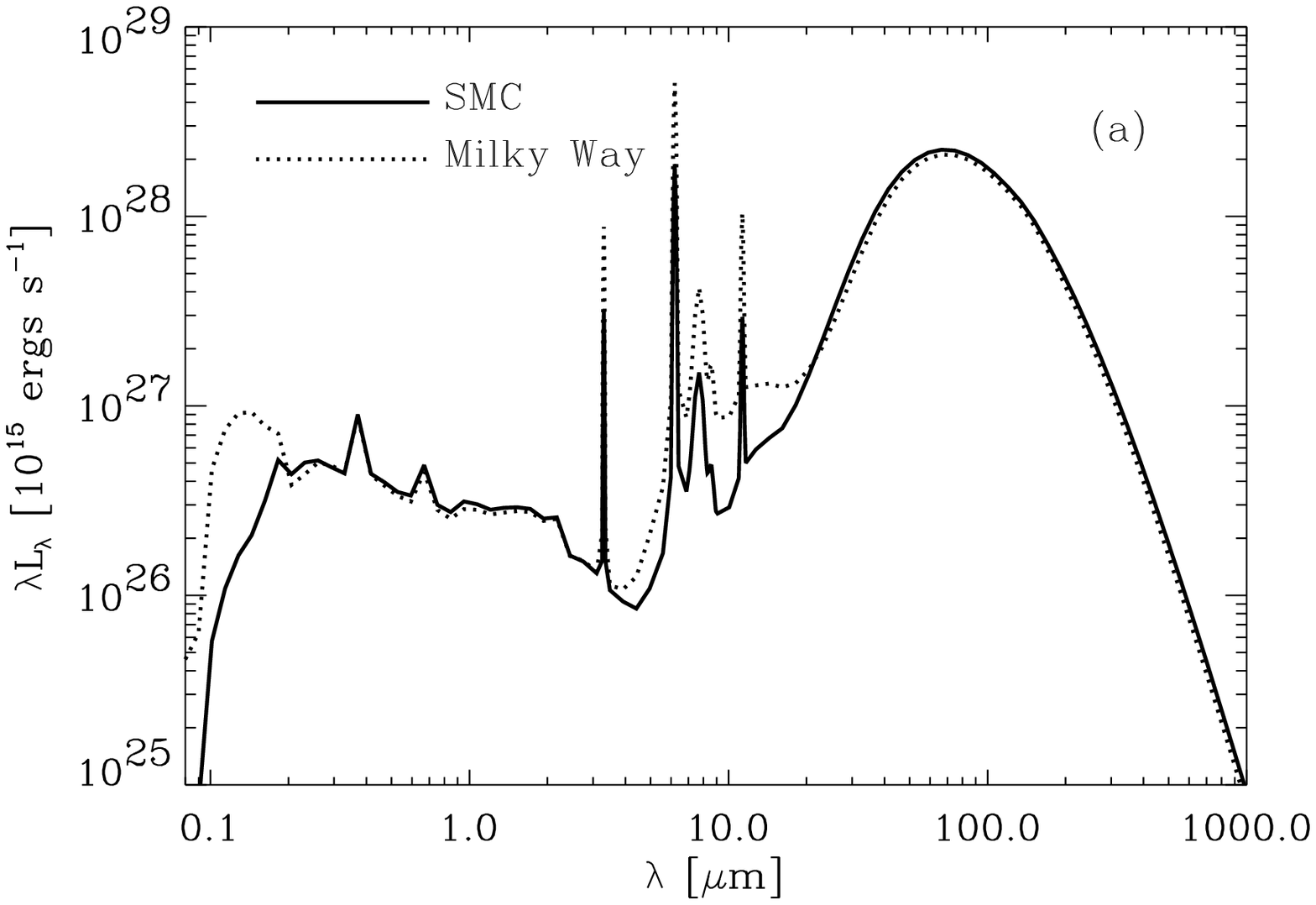}{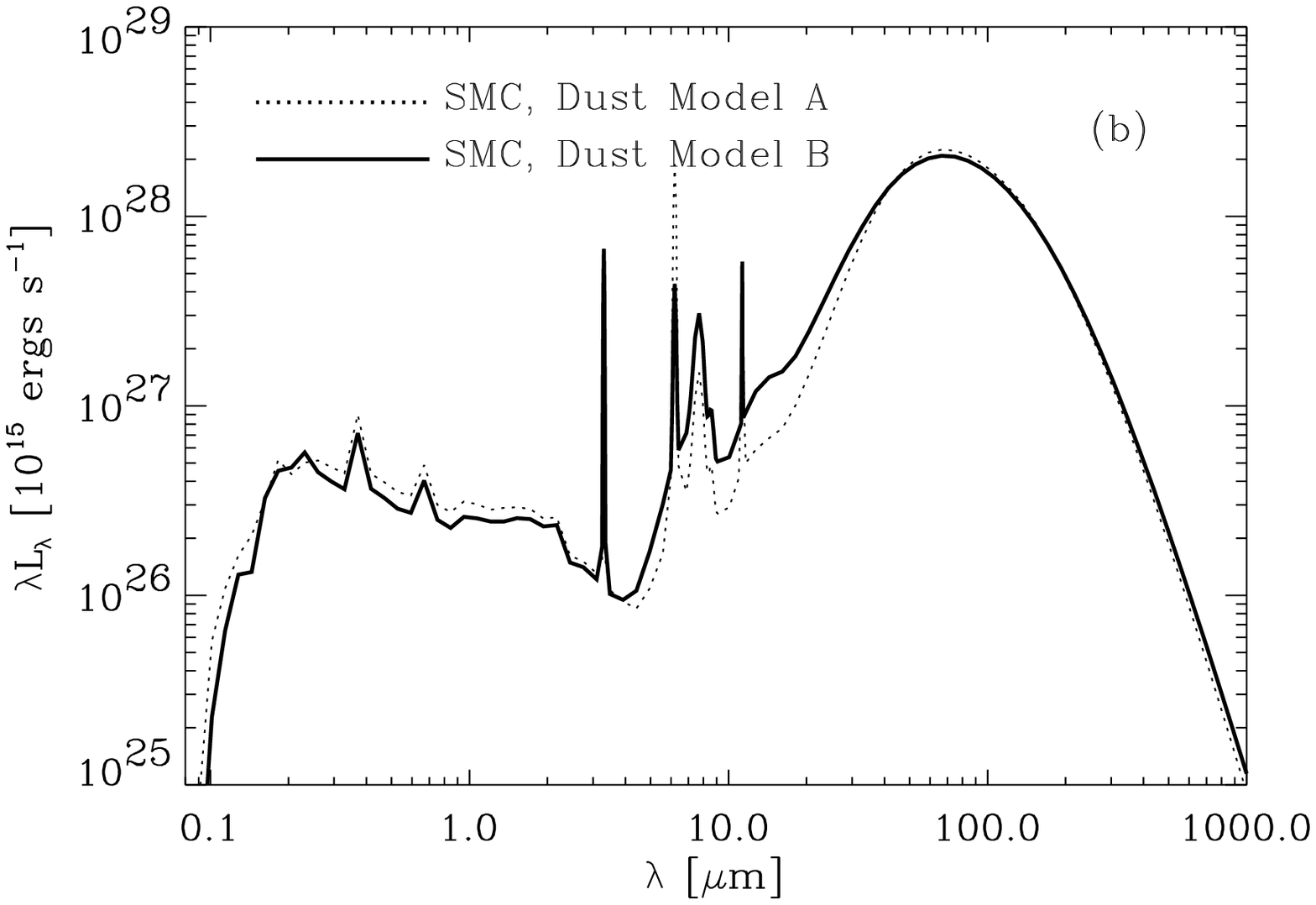}
\caption{Comparison of the effects of using different grain models
  (\S\ref{sec:dust_model}) on the predicted starburst SED. (a)
  Comparison of MW and SMC type dust using grain model A. (b) Comparison
  of SMC type dust using grain models A \& B. All SEDs calculated for same
  set of input parameters (see Table \ref{tbl:model_param}, column 4)
  save the dust type and grain model.
  \label{fig:sed_mwsmc_comp}}
\epsscale{1}
\end{figure}

Observations of UV extinction along different lines of sight in the local group of
galaxies (e.g., mainly the MW, LMC, and SMC) have revealed a range of characteristic
extinction curves which can be broadly associated with the galaxy being observed, although
variations along different lines of sight within a given galaxy can be substantial (e.g.,
Cardelli et al., 1989; Gordon \& Clayton, 1999; Misselt et al., 1999; Clayton et al.,
2000b).  For illustrative purposes, in this paper we concentrate on MW and SMC type dust.
The MW type dust extinction is characterized in the UV by the 2175~\AA\ bump and rising
far UV extinction.  On the other hand, the UV extinction in the SMC is conspicuous in the
absence of the 2175~\AA\ feature.  In addition, the far UV rise in the SMC extinction
curve is nearly linear with $\lambda^{-1}$ and is steeper than in the MW. As discussed in
\S\ref{sec:dust_model}, these extinction curve characteristics are reproduced in our model
by varying the grain size distributions and the relative contributions of the various
grain components to the extinction curve.  Hence, the steepness of the far UV extinction
in the SMC and the absence of the 2175~\AA\ feature require a larger number of small
silicate grains and fewer small graphite grains, respectively, in the SMC dust model
compared to the MW.  In Figure \ref{fig:sed_mwsmc_comp}(a), the difference between
utilizing MW and SMC type dust in our model is illustrated. The difference is manifested
in the UV SED in the presence of an absorption feature near 0.22~$\micron$ in the SED
derived using MW type dust and an increase in the far UV absorption in the SED derived
using SMC type dust.  There are also pronounced differences in the predicted IR SED
depending on the dust model used.  A subtle difference is seen in the depth of the $\sim
9.7~\micron$ absorption feature, which is deeper in the SMC SED compared to the MW curve.
Since the $9.7~\micron$ feature is attributed to stretching and bending resonances in the
small silicate grains \citep{w92}, its strength in the SMC SED is not surprising given
that a large number of small silicate grains are required in the SMC dust model to
reproduce the steep linear rise in the far UV extinction curve.  The most apparent
difference is the excess in mid IR emission present in the MW SED as compared to the SMC.
This is easily understood in terms of the larger populations of small, graphitic grains
and PAH molecules in the MW type dust model.  These grain populations dominate the
emission from grains undergoing transient heating (\S\ref{sec:trans}) and hence there are
more small grains at high temperatures when the MW dust model is used resulting in
increased emission in the mid IR. The mid IR emission from the SMC type dust can be
enhanced by utilizing dust Model B (see \ref{fig:sed_mwsmc_comp}(b)).  The low mid IR
emission from the SMC type dust using dust Model A results from the paucity of small
carbonaceous grains.  By introducing a population of small carbonaceous grains in the form
of AMC, the importance of transient heating is amplified without introducing a 2175~\AA\ 
absorption feature in the resulting extinction curve.

\begin{figure}
  \epsscale{.5}
  \plotone{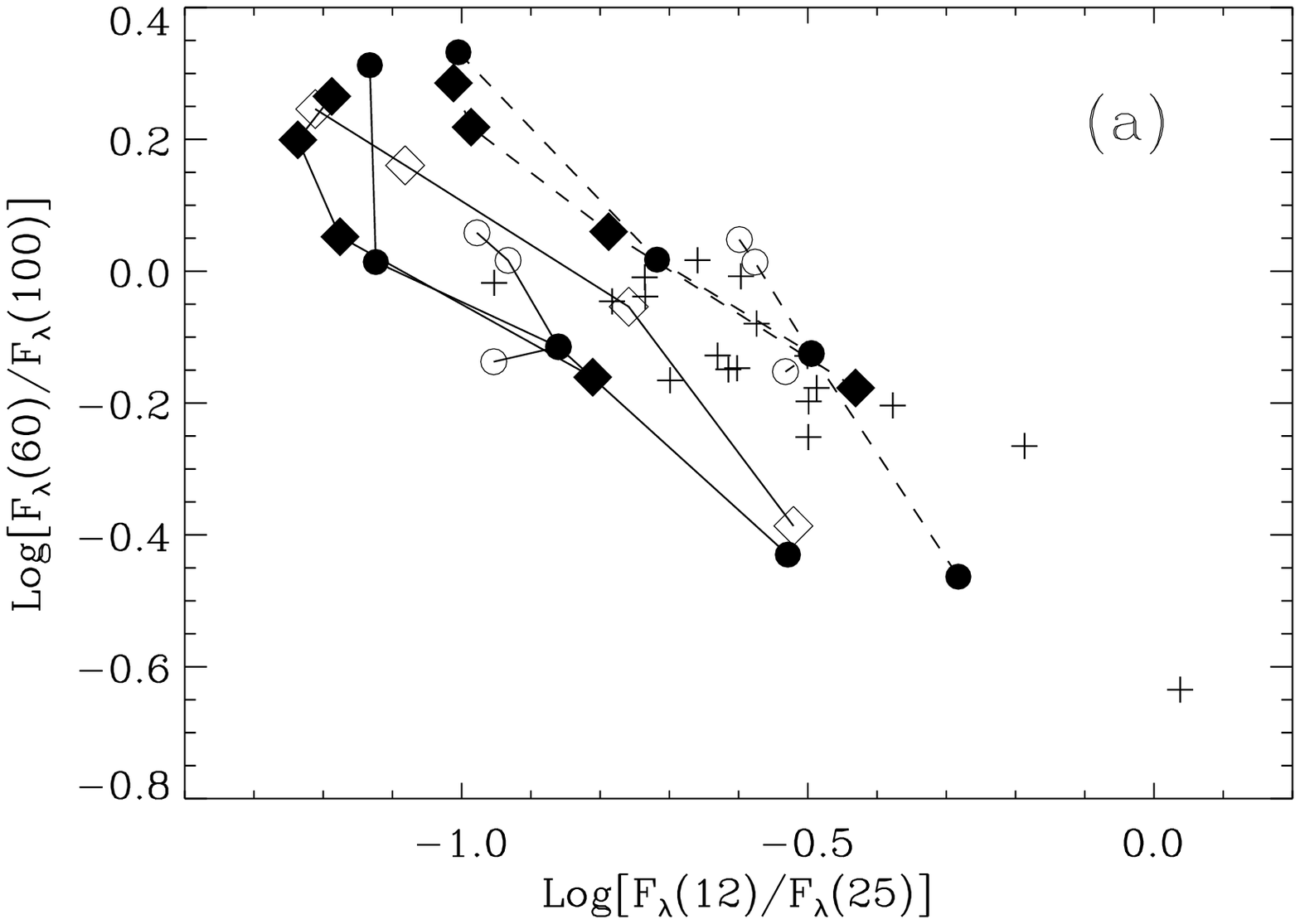}
  \plotone{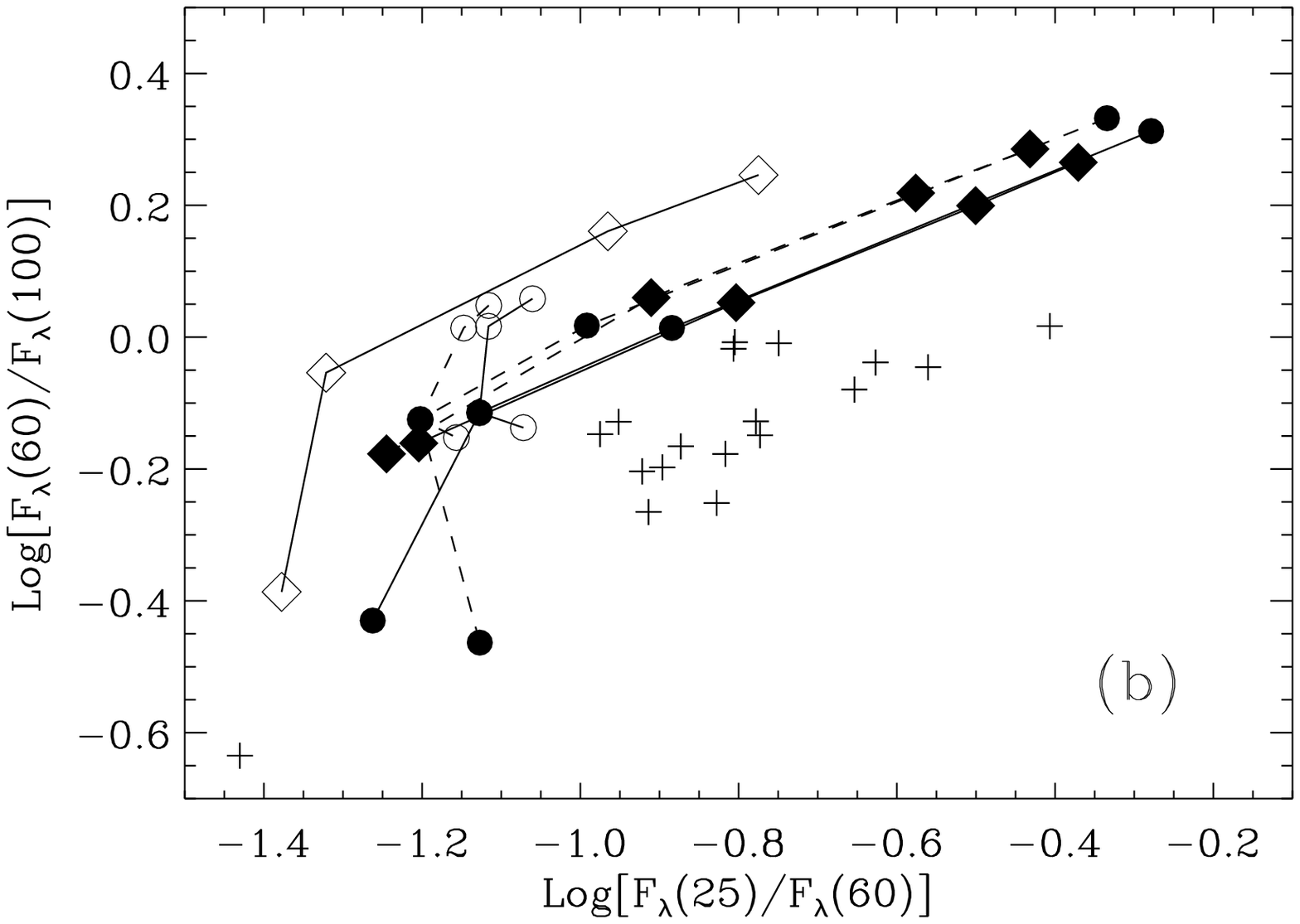}
  \plotone{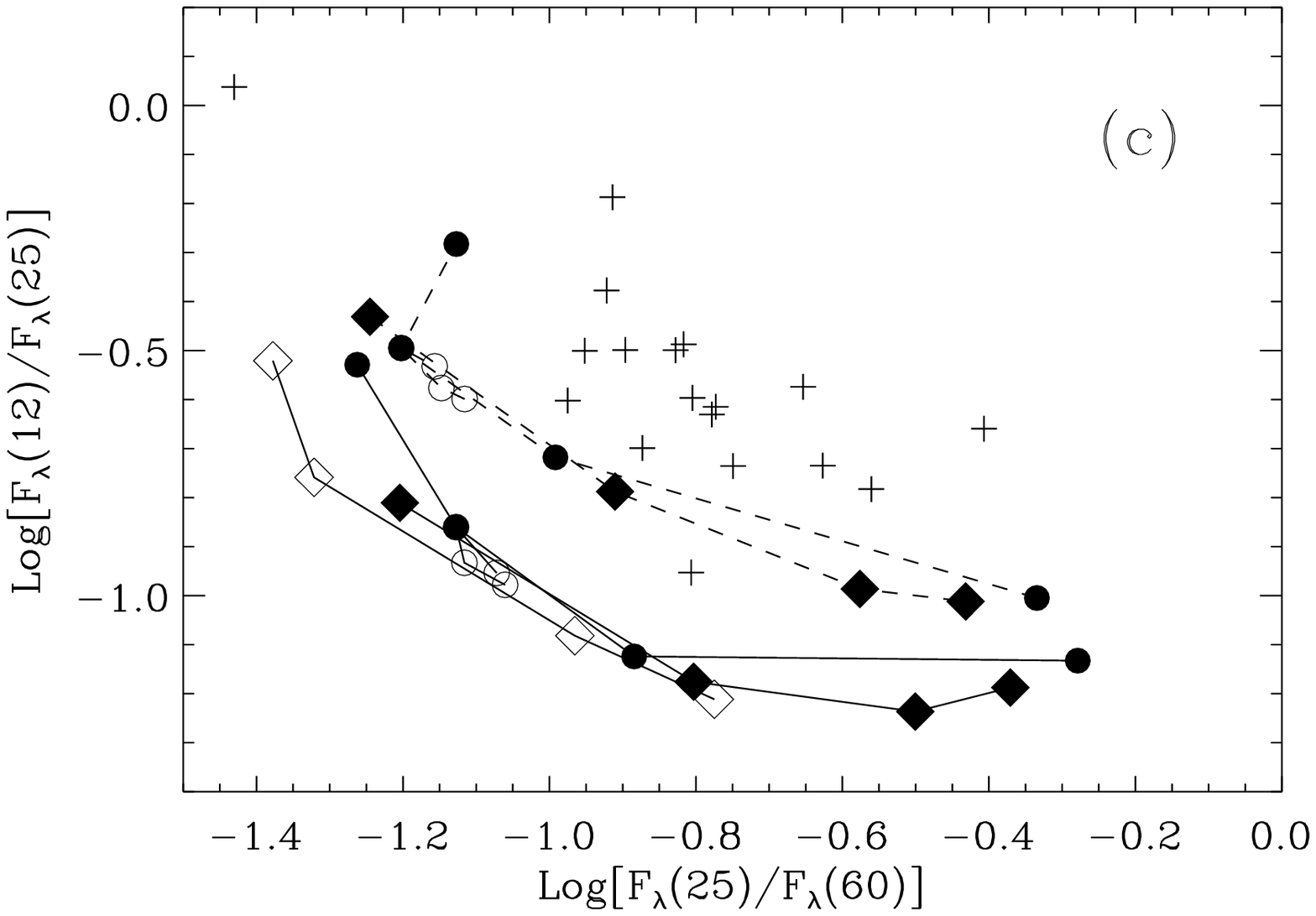}
  \caption{IRAS color--color plot for starburst galaxies with various model runs
    superposed.  The crosses are IRAS data for starburst galaxies from \citet{gcw97} with data in
    all four IRAS bands.  Dashed and solid lines represent models using MW and SMC type dust 
    respectively.  A sequence of varying physical size models (100, 500, 1000, \& 5000~pc)
    is indicated with filled circles.  The open circles represent a series of models with varying
    optical depths with  $\tau_V$=0.5,1.0,10,20.  Sequences of varying SFR models are shown
    with open diamonds for a DUSTY global geometry \& filled diamonds for a SHELL global
    geometry.  Star formation rates along the sequences are 1, 10, 50, \& 100
    M$_\odot$~yr$^{-1}$. All parameters not varying along a sequence are kept fixed at the values
    indicated in column 4 of Table \ref{tbl:model_param}. 
    \label{fig:cc_plots}}
  \epsscale{1}
\end{figure}

Observationally, a substantial population of small grains undergoing temperature
fluctuations would be manifested in higher mid IR fluxes than expected from large grains
in equilibrium.  Indeed, ground based observations \citep{a78,swd83,s84,s+85} along with
early results from the IRAS satellite (e.g., Boulanger et.\ al.\ 1988; Boulanger \&
P\'{e}rault 1988 and references therein) of significant emission in the mid IR in a
variety of environments were a large driving force in the development of ways to treat the
heating of small grains and the recognition that small grains must be a significant
component of interstellar dust \citep{lp84,da85,d86}.  The effect of increased emission in
the mid IR on, for example, IRAS colors, is to increase the
F$_{\lambda}(12\micron)$/F$_{\lambda}(25\micron)$ and
F$_{\lambda}(25\micron)$/F$_{\lambda}(60\micron)$ flux ratios.  As we are illustrating the
behavior of the \dm model in the context of a starburst galaxy model in this paper, in
Figures \ref{fig:cc_plots}a--c, we plot the IRAS colors of starburst galaxies with
measured fluxes in all four IRAS bands from the sample of \citet{gcw97}, along with tracks
from runs of the \dm model.  Runs are included for a range of optical depths, physical
sizes, star formation rates, geometries, and all were run using dust Model A.  Model data
points were determined by convolving model SEDs with the response functions of the IRAS
bandpasses as tabulated in the online IRAS Explanatory Supplement.

The first thing to notice is that the model results cover essentially the full range of
observed starburst colors.  However, in detail, the model colors show some discrepancies
compared to observations, as is especially evident in Figures \ref{fig:cc_plots}b \& c.
The discrepancy between the model colors and the starburst data is attributable to low mid
IR fluxes predicted by the former, especially at 25~\micron.  The low predicted mid IR
fluxes can be traced directly to the dust model.  As can be seen in Figures
\ref{fig:cc_plots}b \& c, the discrepancy between \dm colors and the data is lessened when
the MW dust model is used.  This results from the inclusion of more small graphite grains
and PAH molecules in the MW dust model (\S\ref{sec:dust_model}), which increases the
contribution of transient heating to the mid IR emission (see \S\ref{sec:dis_trans} and
Figures \ref{fig:tr_onoff} \& \ref{fig:sed_mwsmc_comp}).  While it is not our intention
with this paper to explore in detail the wide range of systems to which the \dm model can
be applied, including starbursts, these figures are indicative of the diagnostic potential
of properly done, self--consistent UV to far IR radiative transfer simulations.  The
apparent deficit of small grains could be alleviated by including a separate, large
population of small graphitic grains in the dust model.  However, the absence of a
significant 2175~\AA\ absorption feature in the UV SEDs of many starburst galaxies
\citep{gcw97} makes such a modification of the grain model problematic as the small
graphite grains are responsible for this feature in our dust model.  Hence,
self--consistently reproducing the SEDs of starburst galaxies over a wide wavelength
regime may require more complicated grain models, such as a four component model including
amorphous carbonaceous grains in addition to the PAH, graphite and silicates (i.e., dust
grain model B).  Indeed, using our dust grain Model B with a size distribution appropriate
for the SMC brings the colors of the predicted SED into better agreement with the observed
colors (see Figure \ref{fig:sed_mwsmc_comp}(b)).  Such a diagnostic of grain materials may
provide insight into grain processing histories and grain evolution in response to a wide
range of environmental factors (e.g. Gordon \& Clayton, 1998; Misselt et al., 1999;
Clayton et al., 2000a,b).  Alternatively, in the case of starbursts, the simplistic
assumptions of a single stellar population and relatively simple global geometries in the
models discussed above will likely need to be modified and more complicated arrangements
considered.

\subsection{Global Geometry, $\tau _V$, \& Physical Size}
\label{sec:dis_optdep}
\begin{figure}
  \plotone{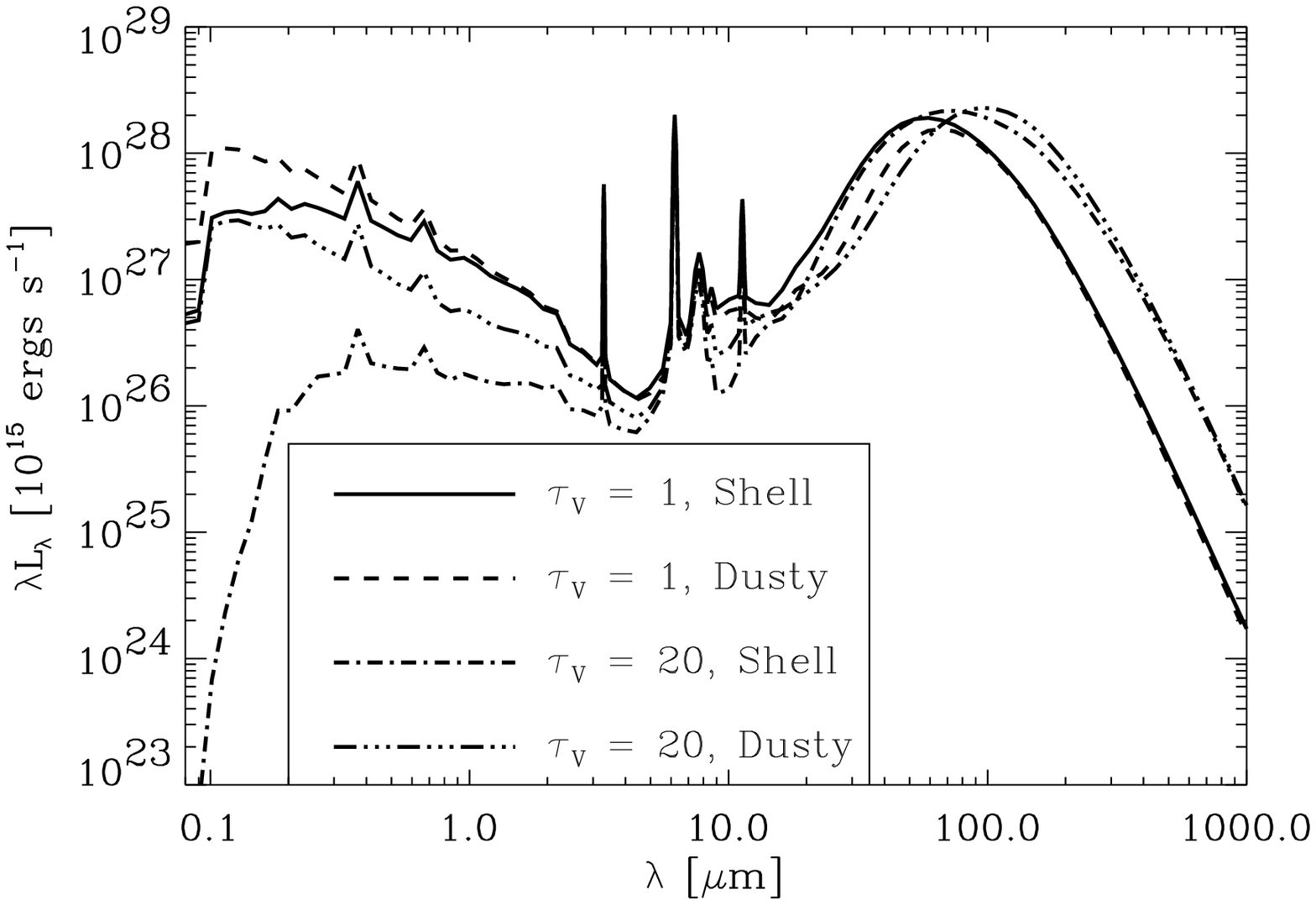}
  \caption{Direct comparison of SHELL and DUSTY geometries at $\tau_V  =1$ and
    $\tau_V=20$. All other input parameters are fixed at
    the values in column 4 of Table \ref{tbl:model_param}.
    \label{fig:sed_var_tv_cgg}}
\end{figure}

\begin{figure}
  \plottwo{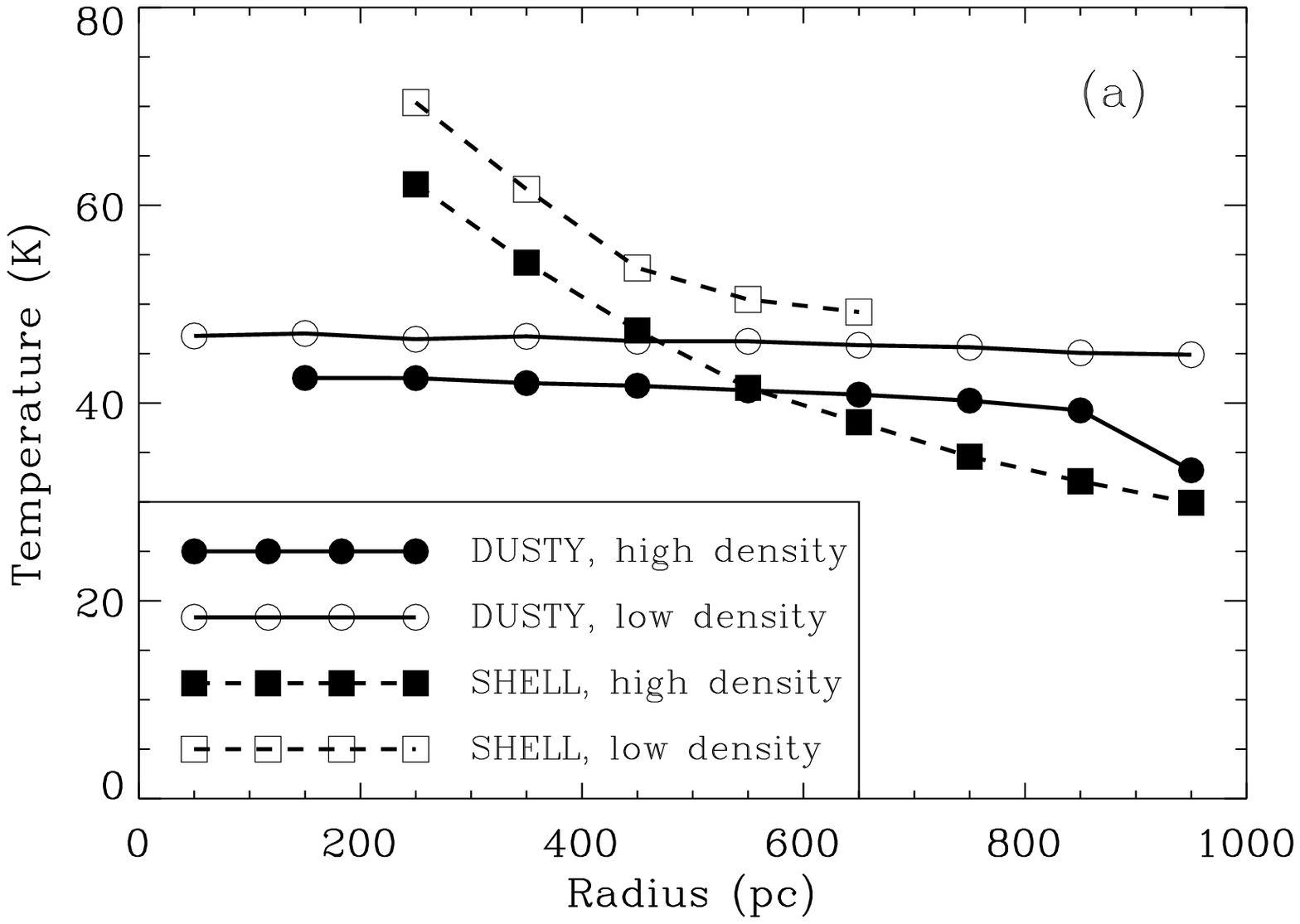}{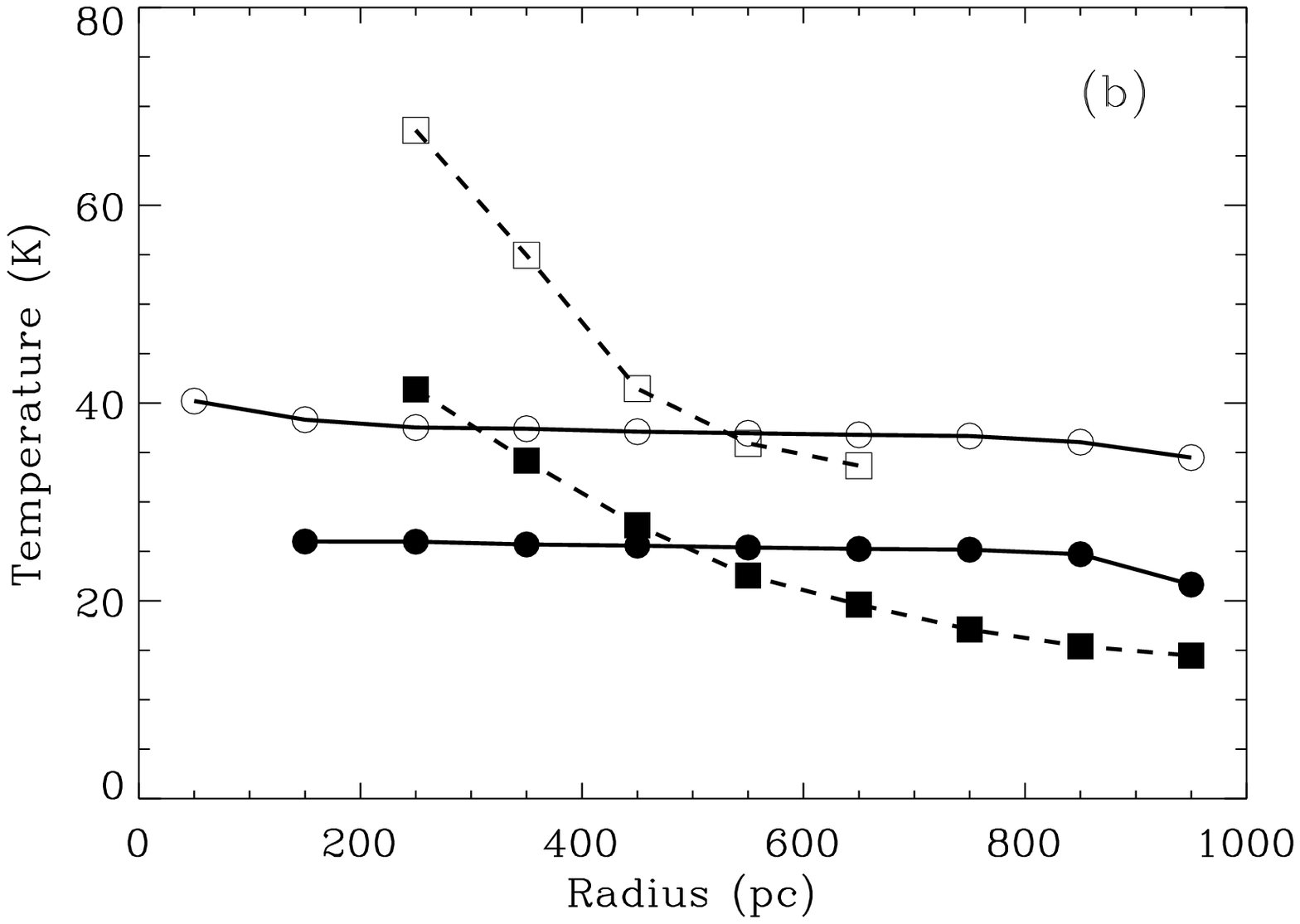}
\caption{Comparison of the radial dust temperature (Eq. \ref{eq:dust_temperature})
  distribution of graphite grains in SHELL and DUSTY geometries for (a) $\tau_V  =1$ and
  (b) $\tau_V=20$. All other input parameters are fixed at
  the values in column 4 of Table \ref{tbl:model_param} in both figures.
  \label{fig:temp_var_tv_cgg}}
\end{figure}

Figure \ref{fig:sed_var_tv_cgg} contrasts the behavior of the SHELL and DUSTY geometries
at the same optical depth.  We plot the model SEDs for both geometries for two extreme
optical depths, $\tau_V = 1$ and $20$ with all other input parameters set to the values in
column 4 of Table \ref{tbl:model_param}.  At a given $\tau_V$, the SHELL geometry absorbs
more energy than the DUSTY geometry since in the SHELL configuration the stars lie inside
all the dust while in the DUSTY configuration the stars are mixed throughout the dust so
energy from stars in the outer parts of the model has a greater chance of escaping the
model space without being absorbed \citep{wg00}.  As a result, at a given optical depth,
the IR emission from the SHELL geometry will always peak at shorter wavelengths.  In
addition, owing to the temperature structure, the SHELL geometry produces a broader IR
SED.  While the dust near the centrally distributed heating sources reaches higher
temperatures in the SHELL geometry, dust in the outlying regions sees fewer high energy
photons and subsequently is heated to lower temperatures.  Since heating sources are
distributed throughout the model with the DUSTY geometry, the resulting temperature
distribution is much flatter than in the SHELL case (see Fig
\ref{fig:temp_var_tv_cgg}a,b), resulting in a narrower IR SED.

For a given physical model size, specifying the input optical depth, $\tau_V$, is
equivalent to specifying the dust mass. The input optical depth is defined as the radial
optical depth that would result were the dusty medium distributed homogeneously in the
model space.  For a clumpy dust distribution, the optical depth along a given line of
sight may have a range of values from a small fraction of $\tau_V$ to several times
larger, depending on $f\!\!f$ and $k$, and the effective optical depth will be
significantly reduced with respect to the homogeneous case \citep{wg96,wg00}.

\begin{figure}
  \plottwo{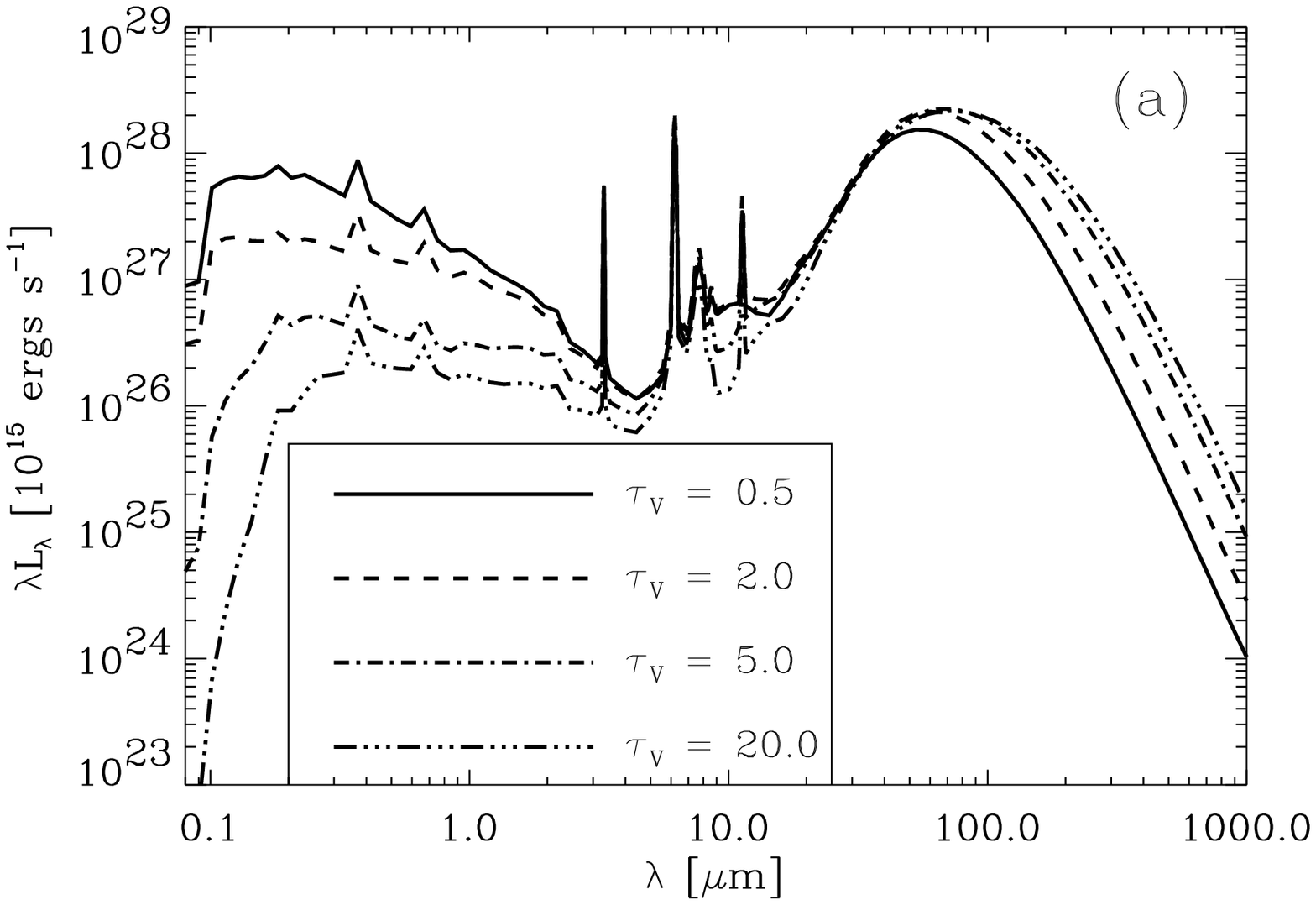}{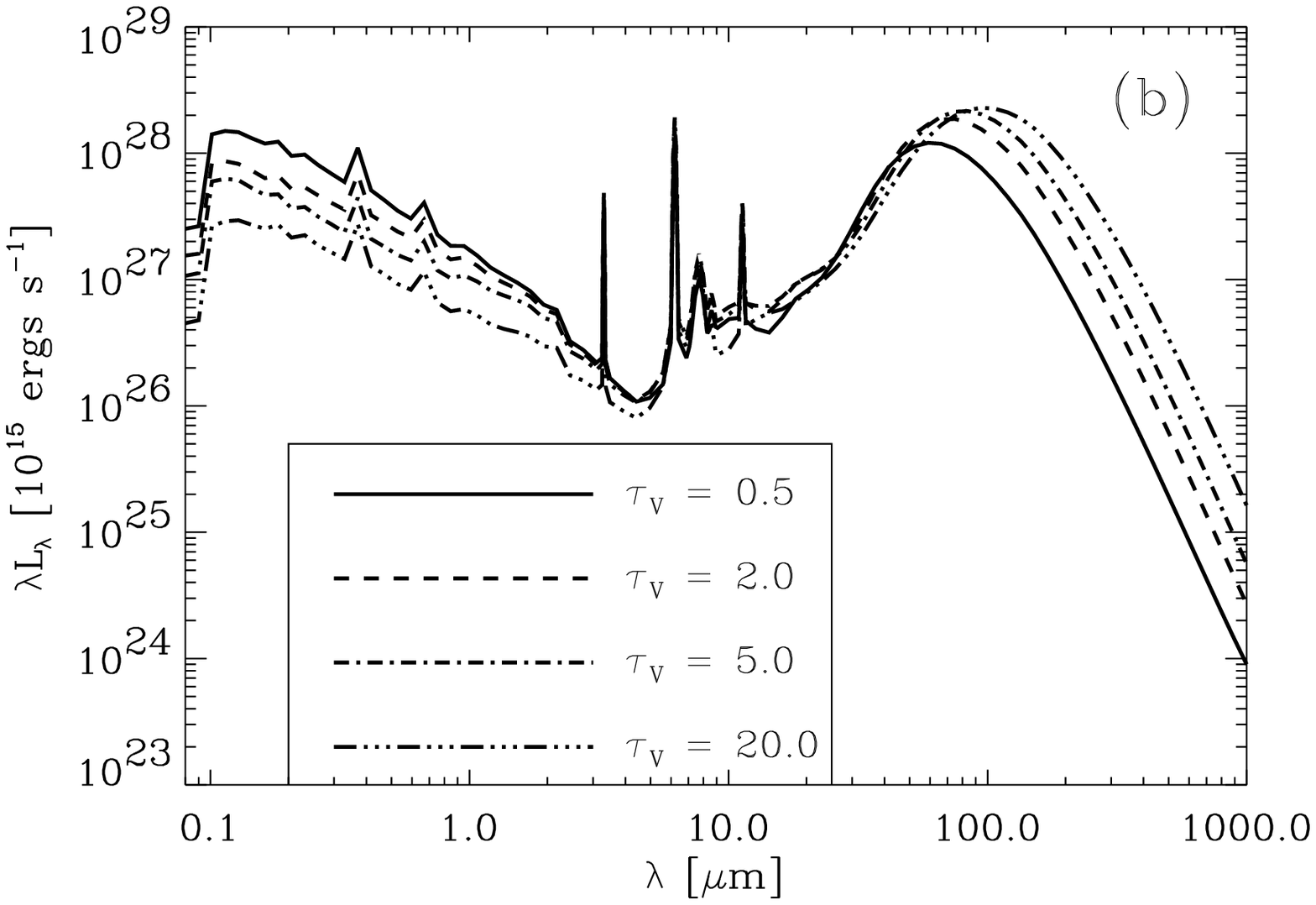}
\caption{Dependence of the model SED on variations in the input optical depth, $\tau_V$,
  for (a) SHELL and (b) DUSTY global geometries.  All other input parameters are fixed at
  the values in column 4 of Table \ref{tbl:model_param}.
  \label{fig:sed_var_tv}}
\end{figure}

\begin{figure}
\plotone{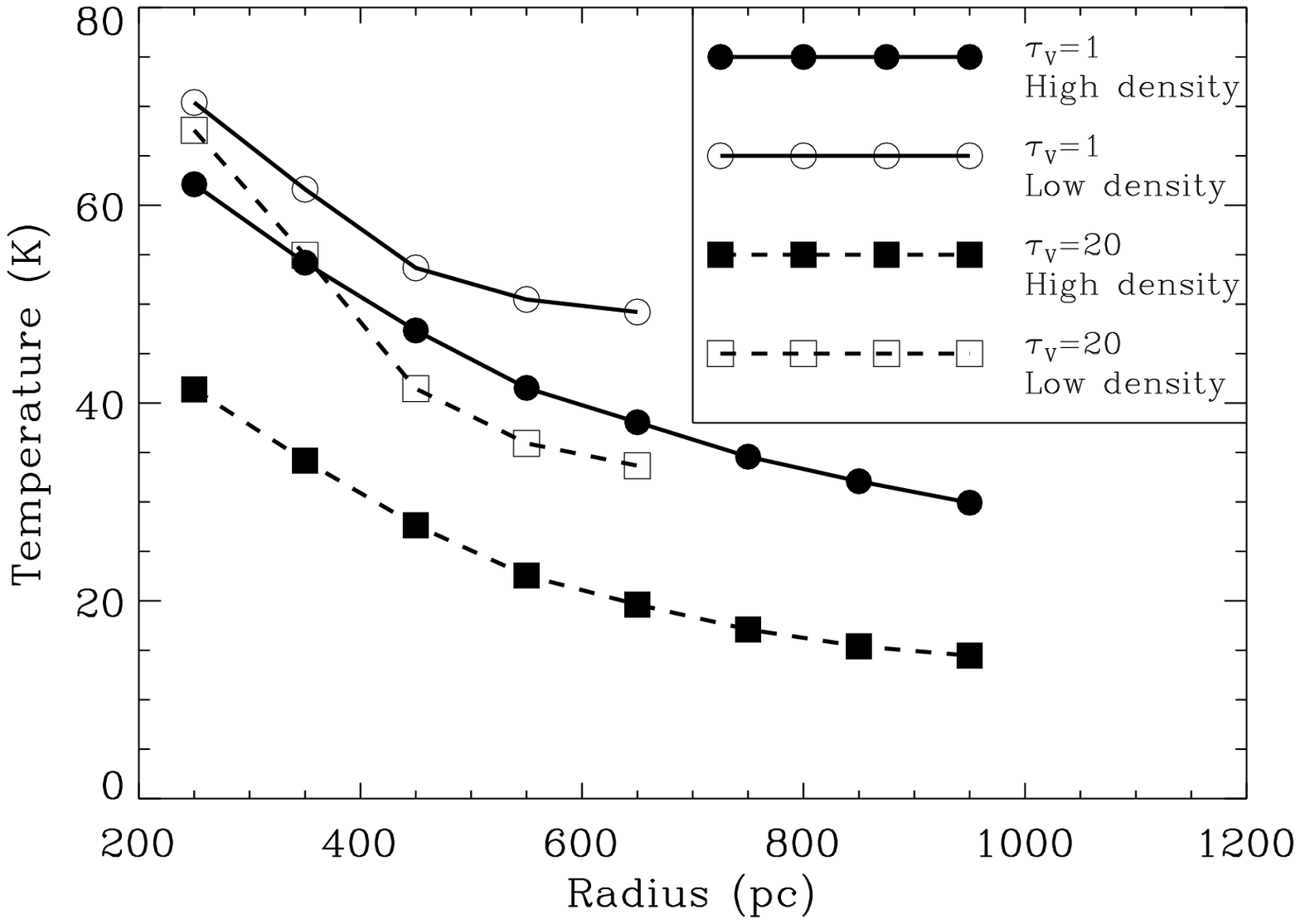}
\caption{Size averaged equilibrium dust temperature (Eq. \ref{eq:dust_temperature}) as a
  function of radial position.  Temperatures in the low and high density model bins
  are plotted for two values of the input optical depth.  For
  clarity, only the temperatures for the graphite grains are shown; results for the
  silicate grain component are similar. All other input parameters are fixed at the values
  in column 4 of Table \ref{tbl:model_param}.
  \label{fig:temp_var_tvc}}
\end{figure}

In Figures \ref{fig:sed_var_tv_cgg} \& \ref{fig:sed_var_tv}a,b, we
show the effects of varying $\tau_V$ on the predicted SED with all other input parameters
kept constant.  Figures \ref{fig:sed_var_tv}a \& \ref{fig:sed_var_tv}b show a series of
model calculations with increasing $\tau_V$ for SHELL and DUSTY geometries, respectively.
In both cases, as $\tau_V$ increases, the IR SED broadens and the peak of the IR emission
shifts to longer wavelengths.  The shift to longer wavelengths, corresponding to lower
dust temperatures, is a result of higher dust masses with increasing $\tau_V$.  So even
though more of the input energy is absorbed as $\tau_V$ increases, there is more dust to
heat and the dust is consequently cooler.  The IR SED broadens as a result of the
increasing importance of the lower density medium in absorbing the input energy.  For a
constant $f\!\!f$ and $k$, as $\tau_V$ increases, the fraction of energy absorbed in the
low density medium increases.  Since the low density medium reaches higher temperatures
than the dense clumps, there is a wider range of dust temperatures for high $\tau_V$ as
compared to the low $\tau_V$ models where most of the energy is absorbed in the dense
clumps (see Figure \ref{fig:temp_var_tvc}).  In addition, at high $\tau_V$, the opacity of
the dust is still significant even at mid IR wavelengths and some of the energy emitted
by the dust is re-absorbed and emitted at longer wavelengths, further broadening the IR
SED.

\begin{figure}
  \plottwo{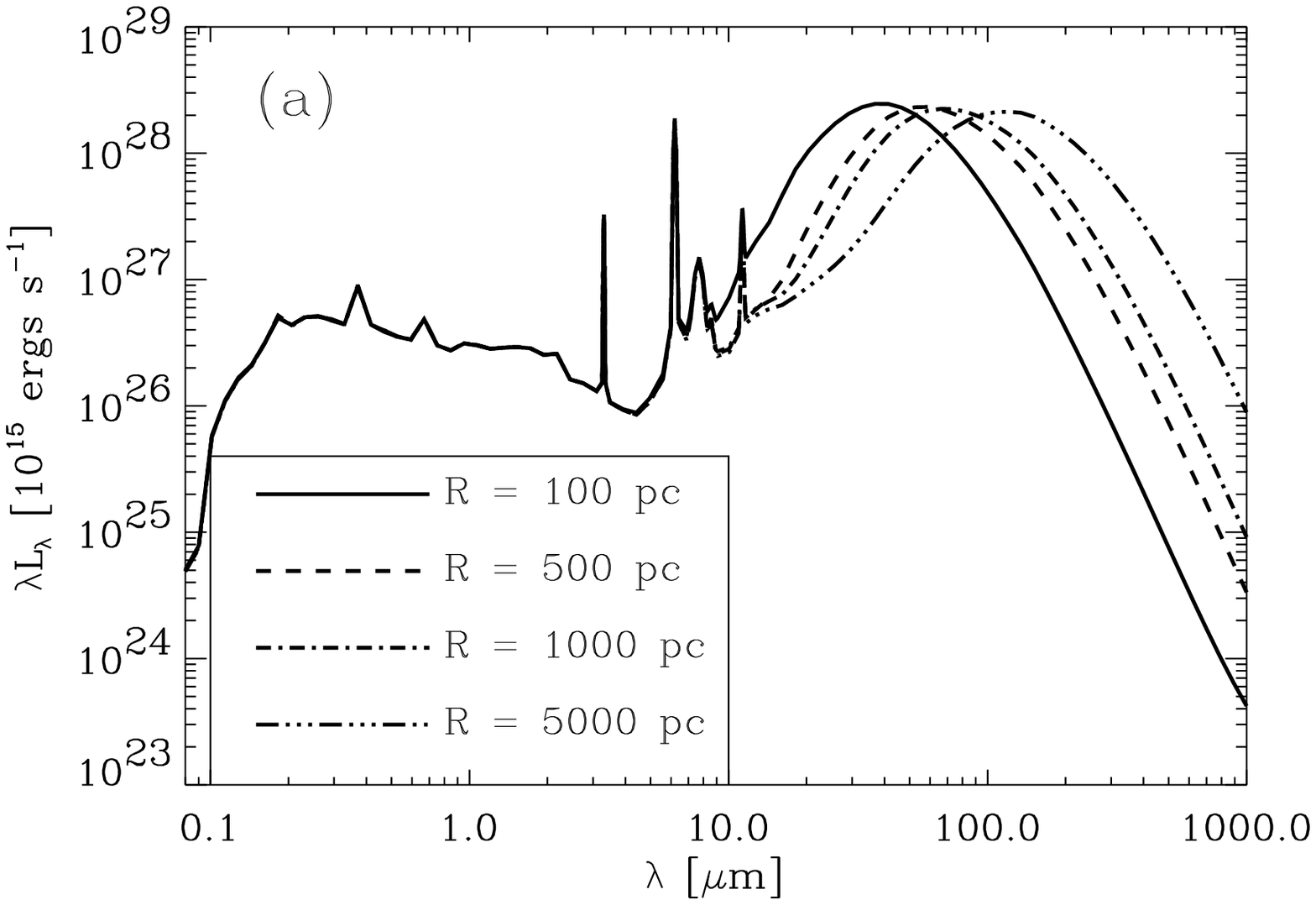}
          {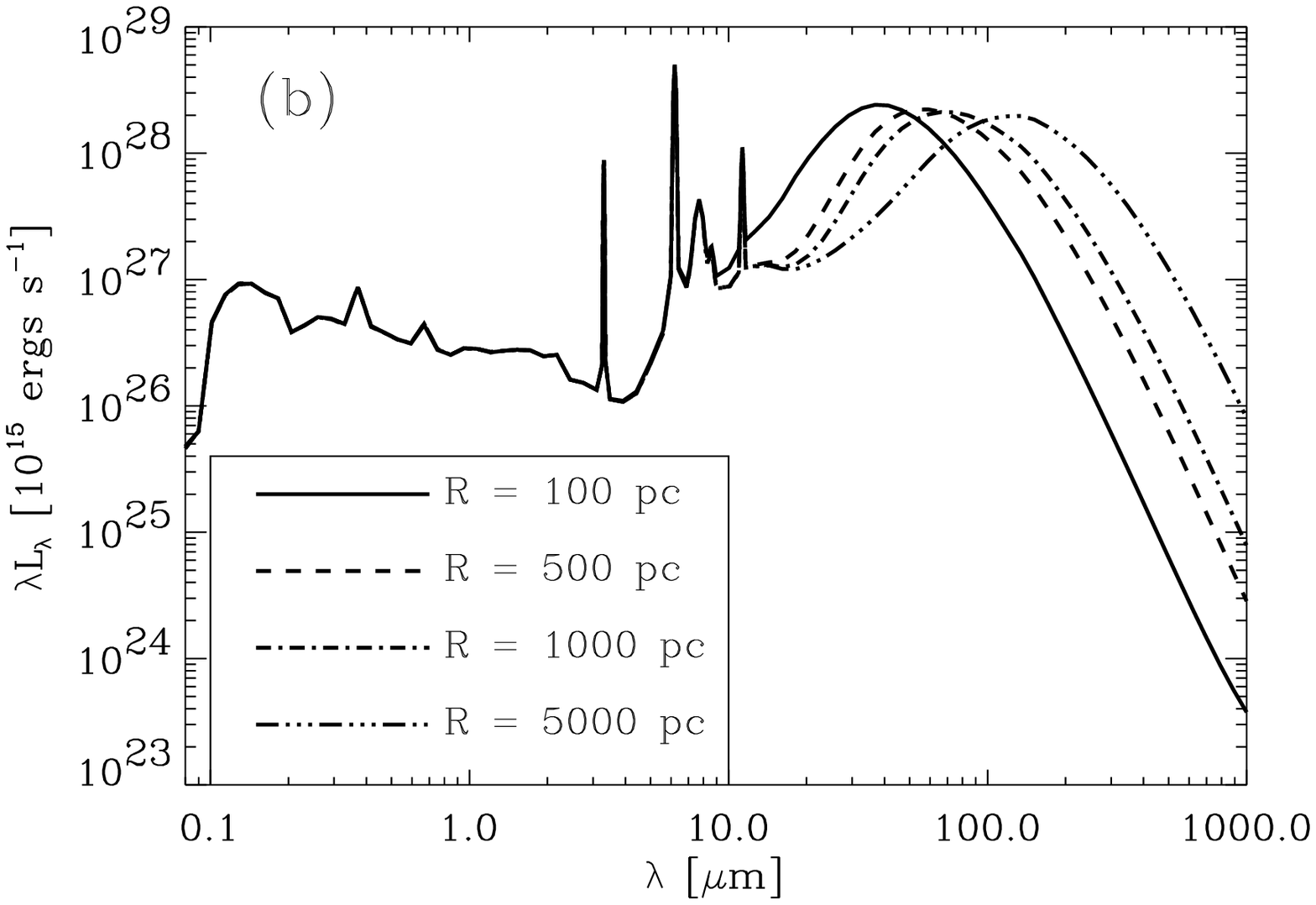}
\caption{Variation of the model SEDs for a range of model sizes for an assumed (a) SMC and
  (b) MW dust model. All other input parameters are fixed at
  the values in column 4 of Table \ref{tbl:model_param}. \label{fig:var_size_sh} }
\end{figure}

Along with $\tau_V$, the physical size of the modeled region will determine the total dust
mass of the system and hence affect the predicted IR spectrum from the dust emission.
There is no dependence in the optical to UV spectrum on the system size since the
radiative transfer (and hence total energy absorbed at any point in the model) depends
only on the total optical depth through a given model bin.  On the other hand, the dust
mass in a given model bin depends not only on the optical depth, but also the physical
size of the bin; the larger the bin, the higher the dust mass (see Eq.
\ref{eq:dust_mass}).  As the temperature an individual dust grain will reach depends on
the ratio of the energy absorbed to the total dust mass, increasing the size of the model
(and hence the size and dust mass of the individual model bins) results in a decrease in
the temperature of individual grains.  Therefore, we expect the peak of the infrared
emission to shift to longer wavelengths as we increase the size of the model.  We see
exactly this dependence in Figures \ref{fig:var_size_sh}a,b where we plot the predicted
SED for a range of model radii from 100~pc to 5000~pc for both (a) SMC and (b) MW type
dust models.  The peak of the IR dust spectrum shifts from $\sim45$\micron\ to
$\sim$200\micron\ over the range of sizes considered.  Of particular importance is the
fact that, although the peak shifts to longer wavelengths, the spectrum is not what would
be observed by simply shifting the smaller radius model SED to longer wavelengths.  The
larger models still exhibit substantial IR emission at shorter wavelengths from
$\sim$10~\micron\ to $\sim$50~\micron, illustrating the importance of the inclusion of
small, transiently heated grains in the model.  The transient heating of the small grains
does not depend on the total energy absorbed.  While the equilibrium heating of the dust
results in lower dust temperatures, the temperature excursions experienced by the small
grains (\S\ref{sec:trans}) result in a contribution to IR SED at shorter wavelengths,
characteristic of hotter dust grains. The same behavior is seen for both dust type models
with the only difference being the higher mid--IR emission from the MW type dust resulting
from a larger contribution from small graphite grains to the SED (\S\ref{sec:dis_dust}).

\begin{figure}
\plottwo{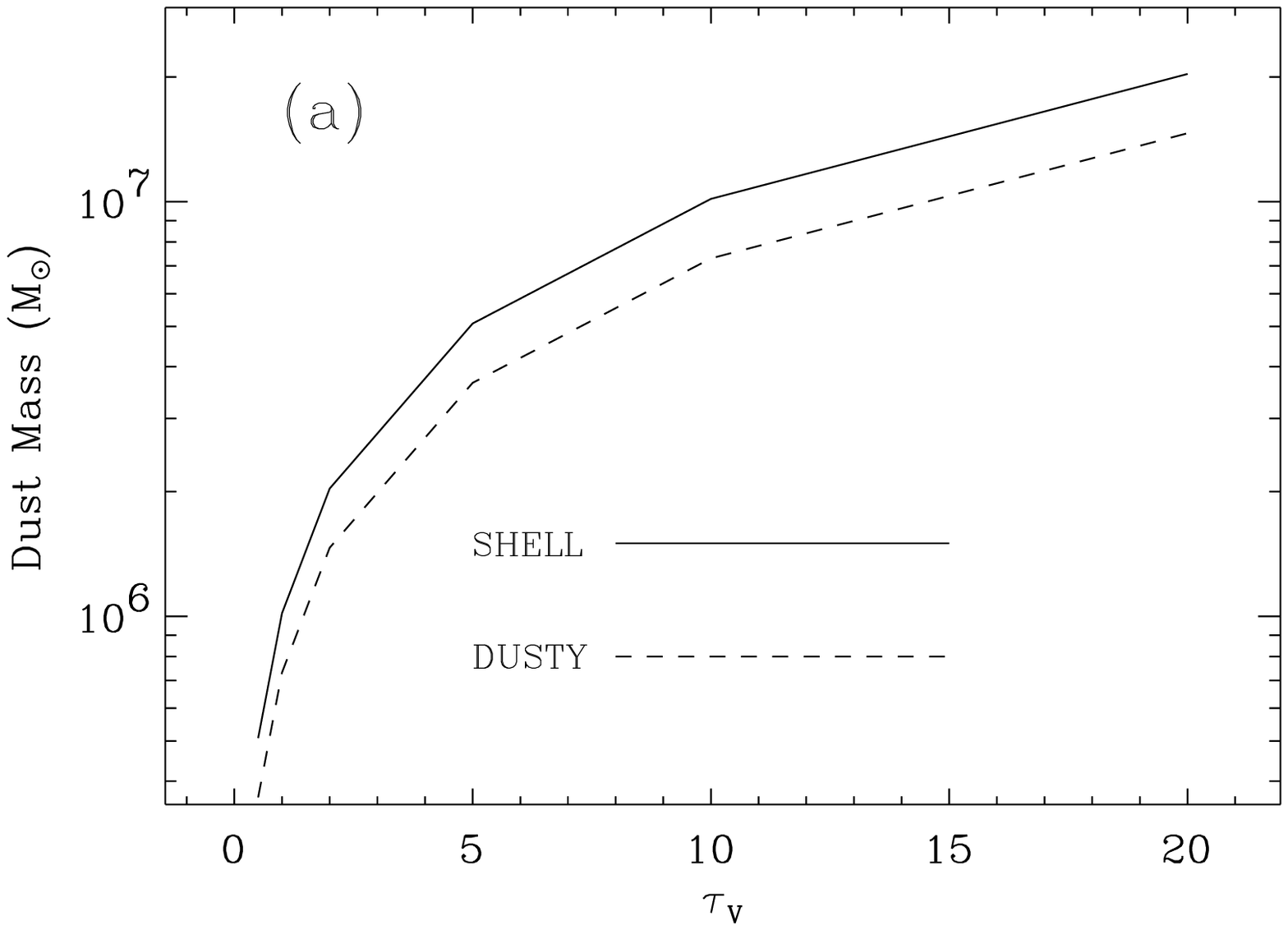}
        {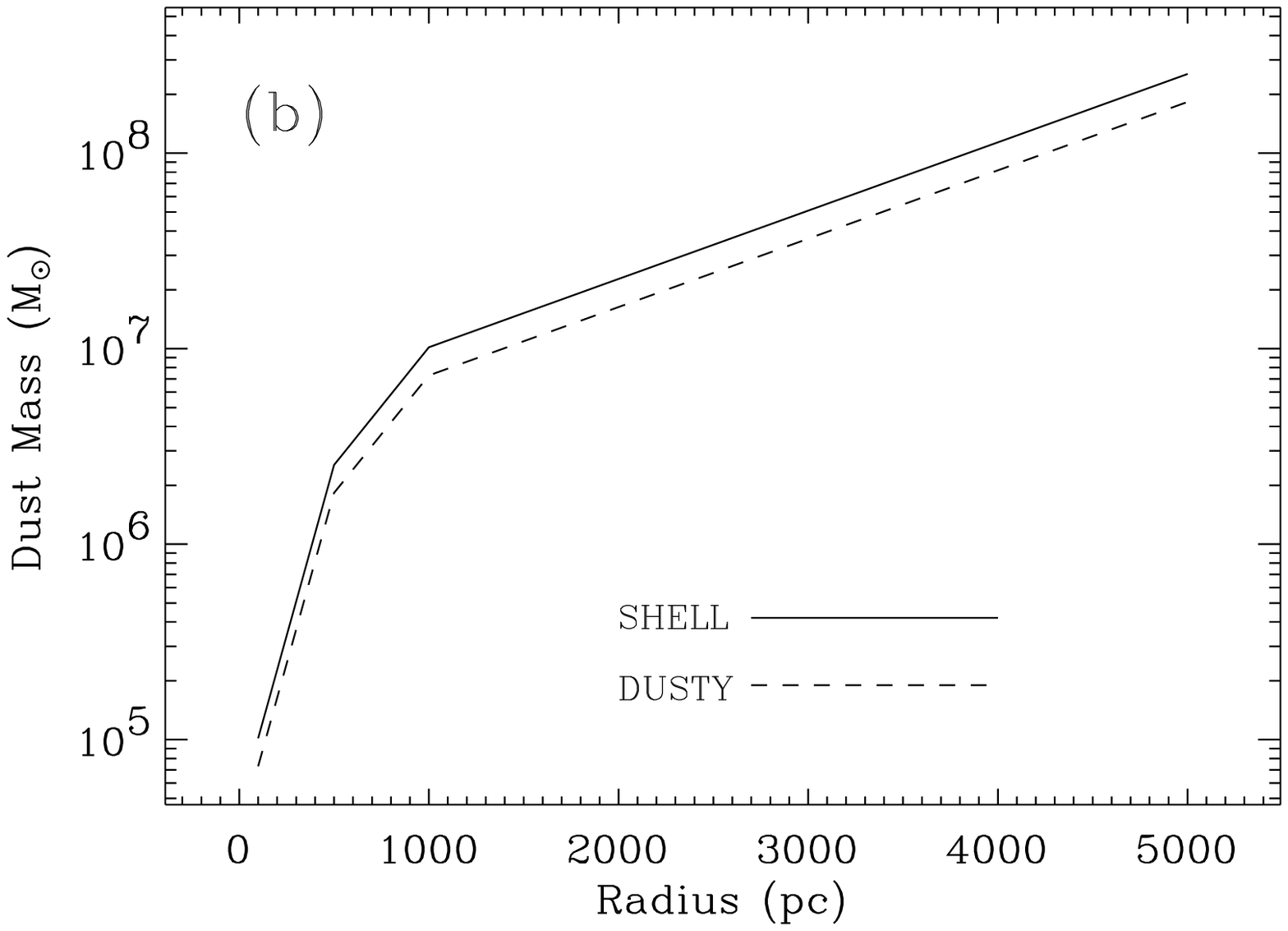}
\caption{Dependence of the total model dust mass on (a) the physical size and (b) $\tau_V$ 
  for both DUSTY and SHELL geometries.  In (a) $\tau_V$ is kept constant at 10 while in
  (b), the physical size is kept constant at 1000~pc.  All other model parameters are
  fixed at the values in Column 4 of Table \ref{tbl:model_param}. 
  \label{fig:dust_mass}}
\end{figure}

The input parameters that determine the dust mass in a given model are the physical size,
$\tau_V$, the dust grain model (MW, LMC, SMC), and the global geometry.  The clumpiness of
the local dust distribution leads to statistical fluctuations in the total mass, but on
average the mass of models with different $f\!\!f$ and $k$ remains the same \citep{wg96}.
The remaining input parameters pertain to the input stellar population and do not affect
the dust mass.  In Figure \ref{fig:dust_mass}, we plot the dependence of the total model
dust mass on (a) $\tau_V$ and (b) physical size for the SHELL and DUSTY geometries.

\subsection{Age and Star Formation Rate}
\label{sec:dis_age} 
\begin{figure}
\plotone{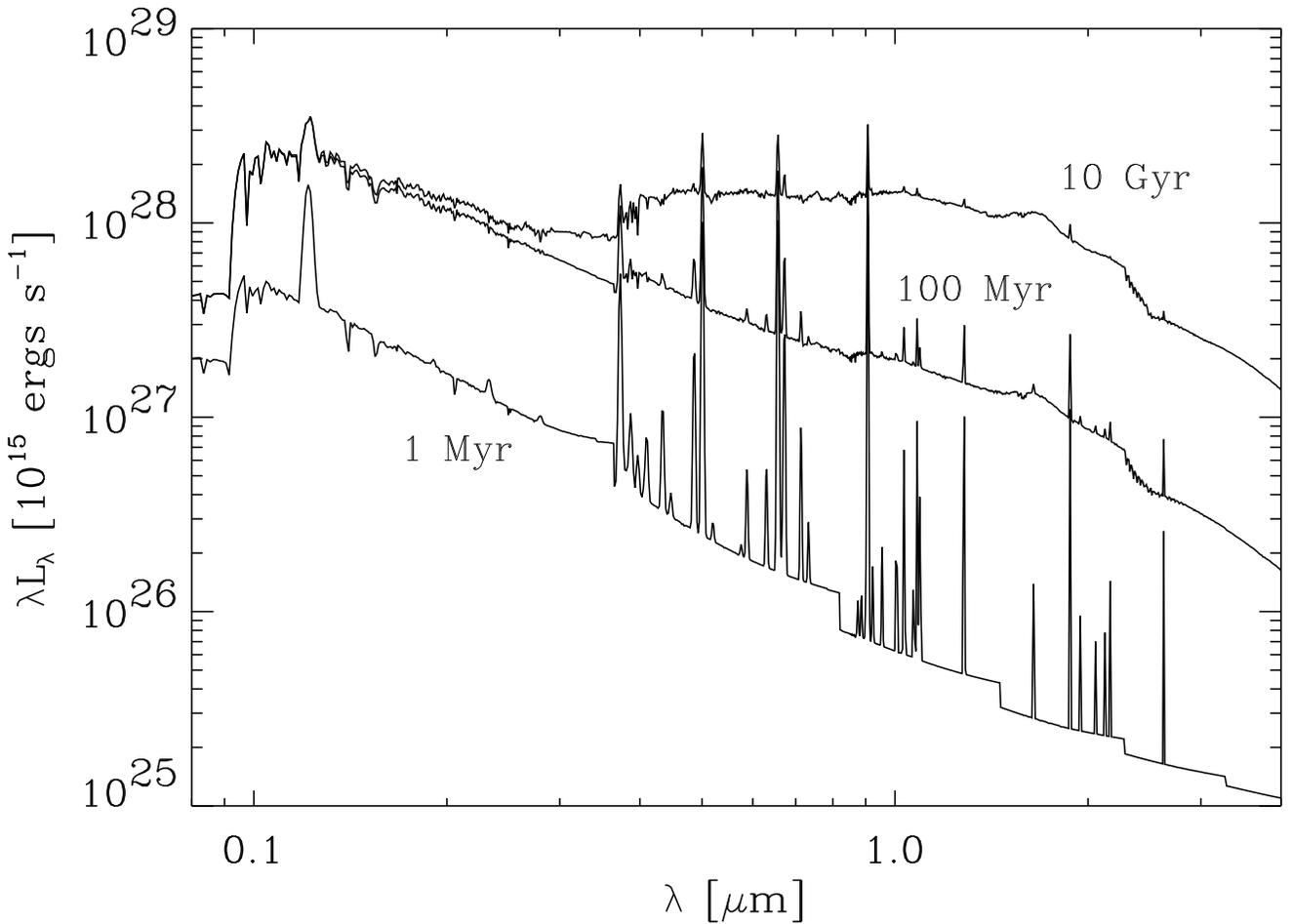}
\caption{Sample input SEDs from the PEGASE synthesis models \citep{fr-v97,fr-v00}, for
  increasing age at a constant SFR of 1~M$_\odot$~yr$^{-1}$. The total mass of the starburst is
  increasing from bottom to top. 
  \label{fig:in_sed_comp}}
\end{figure}

\begin{figure}
\plottwo{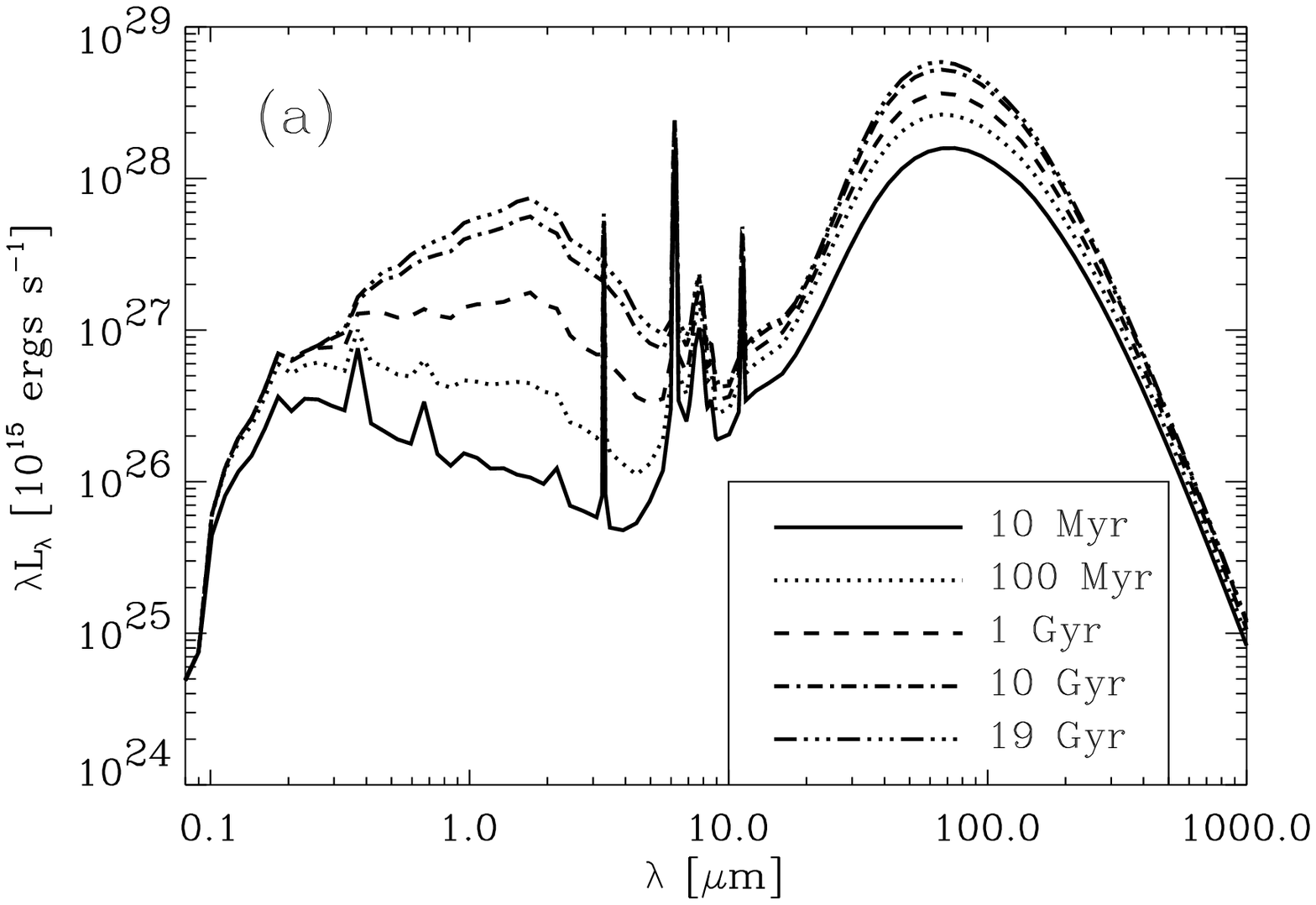}{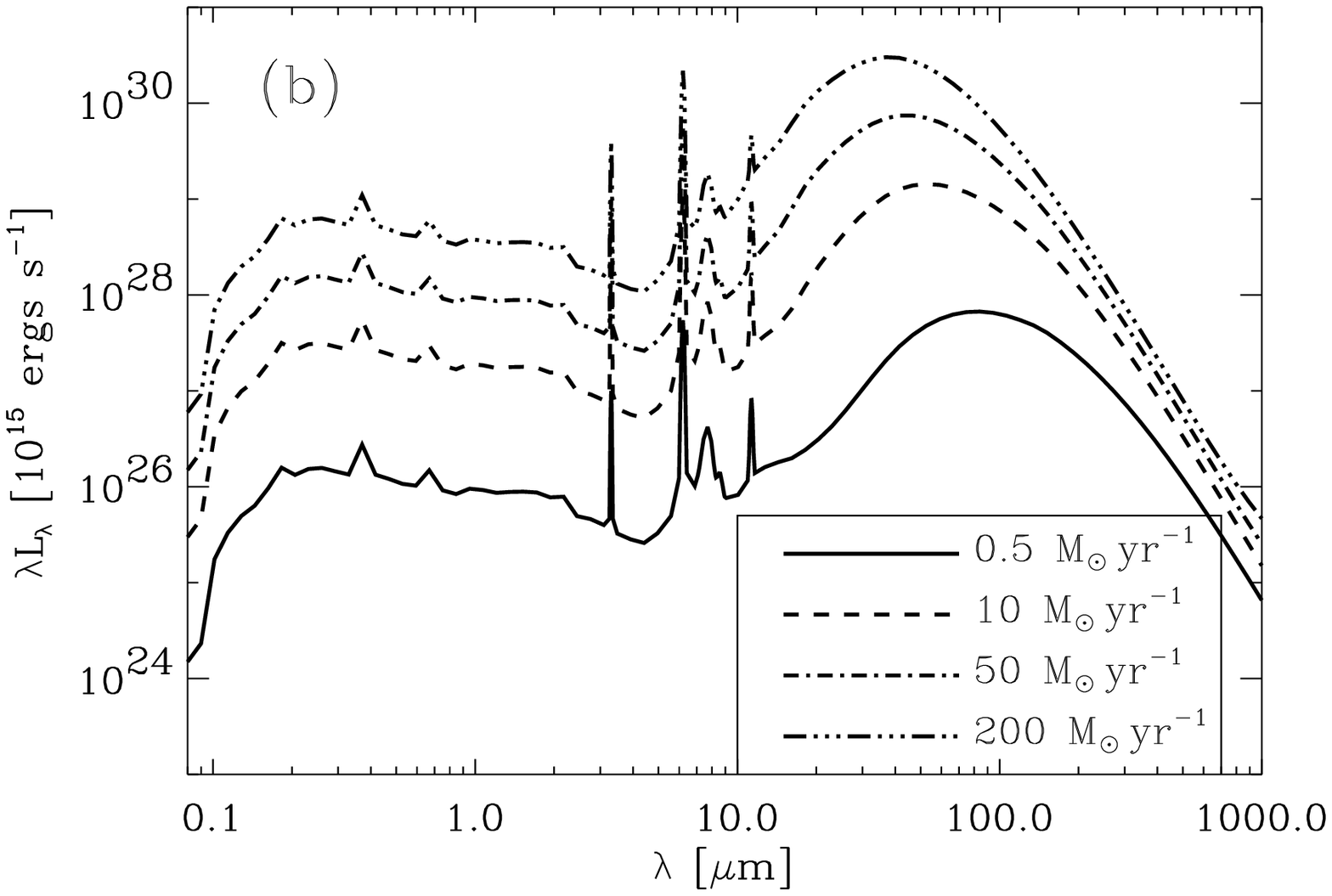}
\caption{Variation of the predicted SED with (a) increasing starburst age for constant SFR 
  of 1.6~M$_\odot$~yr$^{-1}$ and (b) increasing SFR at a constant burst age of 40~Myr.
  All other input model parameters are kept fixed at the values given in column 4 of Table 
  \ref{tbl:model_param}.
  \label{fig:var_age_sfr}}
\end{figure}

\begin{figure}
\plotone{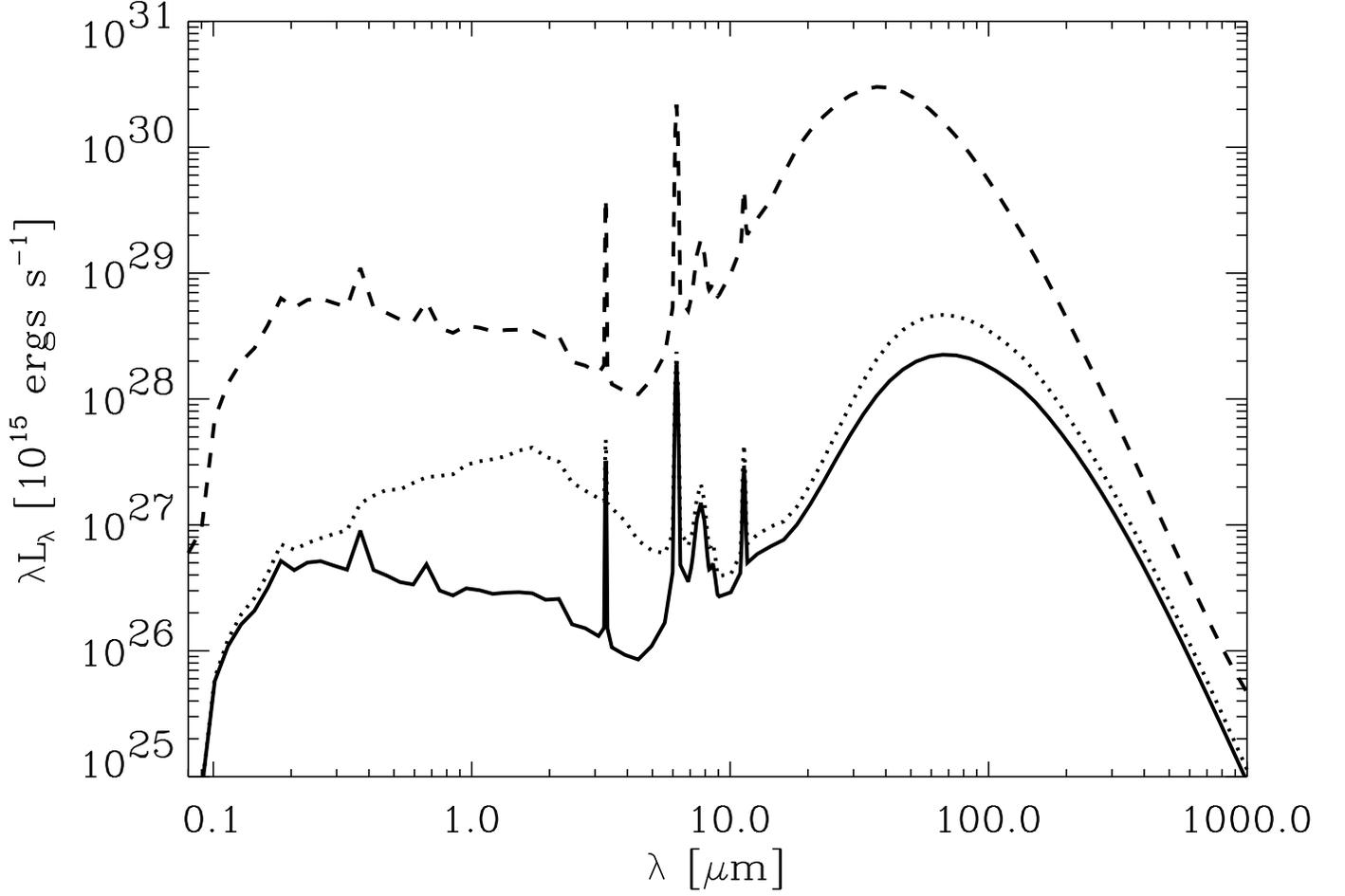}
\caption{Comparison of predicted model SEDs with same total stellar mass. Solid line,
  total stellar mass of $6.4\times10^{7}$~M$_\odot$ with SFR=1.6~M$_\odot$~yr$^{-1}$,
  age=40~Myr.  The dotted line SED results from increasing the total stellar mass to
  $8\times10^{9}$~M$_\odot$ by increasing the starburst age to 5~Gyr while keeping the SFR
  constant at 1.6~M$_\odot$~yr$^{-1}$.  The dashed line SED is the result of increasing
  the total stellar mass to $8\times10^{9}$~M$_\odot$ by increasing the SFR to
  200~M$_\odot$~yr$^{-1}$ while keeping the age constant at 40~Myr. All other input model
  parameters are kept fixed at the values given in column 4 of Table
  \ref{tbl:model_param}.
  \label{fig:cnsmass_comp}}
\end{figure}

The application of the \dm model to starburst galaxies requires the specification of the
stellar content of the galaxy along with its star formation history.  In general, the star
formation history can be characterized by either a burst scenario, wherein all the stars
form at once and subsequently evolve, or by a constant star formation scenario, wherein
stars form continuously as the starburst ages.  While the situation in real starburst
galaxies is likely somewhere in between, for the purpose of illustrating the \dm model, we
consider only the constant star formation scenario.  For this scenario, an increase in
either age or star formation rates will lead to an increase in the total mass of the
starburst.  Increasing the age of the starburst while keeping the SFR constant, while
increasing the total stellar mass, results in an increase of the importance of the
contribution of older, less luminous stars to the SED relative to the hot, young stars
which dominate the UV.  As the starburst ages, the massive young stars are continuously
replenished and their contribution to the UV SED approaches a constant as an equilibrium
is reached between their formation and evolution.  Hence, increasing the mass of the
starburst by increasing its age results in a higher and higher fraction of its energy
being produced in the optical and NIR wavelength regime.  The result is a change in the
shape of the input SED as the starburst ages, the UV approaching a constant with the
optical and NIR increasing in importance; see Figure \ref{fig:in_sed_comp}.  On the other
hand, increasing the mass of the starburst by increasing the SFR at a constant age has the
effect of simply scaling up the input SED.  At a given age, increasing the SFR results in
an increase in the number of old {\em and} young stars and the total emitted energy
increases at all wavelengths. The effects of these variations in the shape and total
luminosity of the input SED on the predicted IR dust emission spectrum are illustrated in
Figures \ref{fig:var_age_sfr}a, \ref{fig:var_age_sfr}b, and \ref{fig:cnsmass_comp}.

As is seen in Figure \ref{fig:var_age_sfr}b, increasing the SFR rate at a constant age
results in a shift of the peak of the IR emission to shorter wavelengths.  Since the
shape of the input SED does not change, but only the total amount of input energy, the
dust absorbs the same fraction of the total energy ($\sim$97\% for the cases plotted here)
and the absorbed energy is distributed between the high and low density clumps in the same
fractions as well.  The result is that as the total amount of input energy is increased,
all of the dust is heated to higher temperatures and the IR emission increases and its
peak shifts to shorter wavelengths.  Similarly, in the case of increasing age, the total
IR emission from dust increases and the peak of the IR emission also shifts to shorter
wavelengths.  However, the effects are much less pronounced (Figure
\ref{fig:var_age_sfr}a).  The reason for the smaller effect is three-fold. (1) Since the
stellar population is aging, the same mass of stars produces less energy compared to a
young, high SFR burst. Therefore, increasing the mass of the starburst by increasing its
age results in a much smaller increase in the energy available for absorption in the dust
as compared to increasing the mass by the same amount by increasing the SFR. (2) The input
SED has changed shape so there is a larger contribution to the total input energy at
optical and NIR wavelengths, resulting in less efficient dust absorption since the
efficiency of dust at absorbing and scattering radiation decreases with increasing
wavelength. As a result, energy is less likely to be absorbed and the fraction of the
total energy absorbed by the dust falls from $\sim$97\% to $\sim$83\% for the cases
plotted in Figure \ref{fig:var_age_sfr}a. (3) Also as a result of the decreasing
efficiency of dust absorption, a higher fraction of the total energy is absorbed
in high density bins with increasing starburst age.  Since the low density medium is
heated to a higher temperature for a given input energy, the shift to increasing
importance of the high density clumps results in a smaller overall temperature increase,
and hence less of a shift in the peak of the dust emission to shorter wavelengths, than
would be the case if the relative fraction of energy absorbed in the high and low density
media remained constant.  For the cases plotted in Figure \ref{fig:var_age_sfr}a, the
fraction of the energy absorbed in the high density clumps increases from $\sim$68\% to
$\sim$82\% for the starburst ages considered.

The difference between the constant SFR rate and the constant age models is illustrated in
Figure \ref{fig:cnsmass_comp}.  The solid line represents the SED of a model with a total
stellar mass of $6.4\times10^{7}$~M$_\odot$ (SFR=1.6~M$_\odot$~yr$^{-1}$, age=40~Myr). We
increase the mass of the model to $8\times10^{9}$~M$_\odot$ by increasing the SFR to
200~M$_\odot$~yr$^{-1}$ keeping the age constant (dashed line) and by keeping the SFR
constant and increasing the age to 5~Gyr (dotted line). Increasing the
age increases the total dust emission by a small fraction and shifts the peak wavelength
of the dust emission slightly to shorter wavelengths.  On the other hand, increasing the
SFR dramatically increases the total dust emission and shifts the peak wavelength
significantly.

\section{Conclusion and Summary}
\label{sec:conclusion}

In this paper and a companion paper \citep{gmwc00}, we have presented the \dm model, a
self-consistent Monte Carlo radiative transfer and dust emission model.  The strengths of
\dm include:

\begin{itemize}
\item Self-consistency.  No {\em ad hoc} assumptions about the dust temperature are
  made; the dust is heated by the absorption of photons originating from sources 
  included in the model.  The temperature distribution of the dust grains, and hence their
  emission spectrum, is calculated self-consistently by keeping track of the energy absorbed
  by dust in the Monte Carlo radiative transfer.
  
\item Treatment of the heating of and emission from small grains, including the
  effects of temperature fluctuations resulting from the absorption of single photons.
  
\item The ability to properly treat high optical depth situations. The iterative solution
  outlined above, wherein dust self-absorption is considered, permits the treatment of
  high optical depth cases, where the optically thin assumption may be violated even at IR
  wavelengths.

\item Properly including the FIR emission from dust in \dm provides additional information
  regarding the dust model, the nature of the heating source(s), and the physical size of the
  modeled region, not available through UV--NIR modeling alone.  
 
\item The use of Monte Carlo techniques to solve the radiative transfer equations allows
  the treatment of arbitrary dust and heating source distributions as well as
  inhomogeneous dust distributions (clumpiness).  Though we have concentrated on starburst
  galaxies in exploring the parameter space of \dm$\!\!$, it is of general applicability and
  can be used to model e.g., dust tori in AGNs and Quasars, circumstellar discs,
  individual star forming regions, and reflection nebulae.
 
\end{itemize}

To emphasize this last point, we point out that the fits to any UV--NIR SED are
degenerate; there are in general several possible combinations of geometries, dust models
and heating sources which can provide essentially identical fits to the SED.  However, by
examining the IR dust spectrum (total energy emitted by dust, peak wavelength of the IR
emission, the strength of the mid IR emission relative to longer IR wavelengths, and
strength of IR absorption and emission features), some of the degeneracy in the UV--NIR
models can be lifted.  Indeed, the FIR dust spectrum provides information on the physical
size of the region not accessible to UV--NIR modeling alone.

Future papers will detail the application of \dm to interpreting observations of
astrophysical systems including individual starburst galaxies.  Improvements to the
current model are also being implemented.  We have presented a
fairly simple treatment of the PAH component here;  we have not included a means of
varying the relative strengths of the PAH features.  Since the strengths of some of the features
depend on the number of hydrogen atoms in the molecule, while others depend on the number
of carbon atoms, the relative strengths can be varied by adjusting the hydrogen coverage,
$x_H$ (see \S\ref{sec:dust_model}).  Applications to specific, individual objects will
require this sort of fine tuning as the relative strengths of the MIR emission features
are observed to change from object to object and environment to environment
\citep{cohen+89}.  A major area of effort will be in applying \dm in a starburst model
that will include multiple stellar populations and evolutionary effects.  This will
require the inclusion of the effects of the interaction between on-going star formation and 
dust, including the evolution of the grains due to processing, formation, and destruction
(e.g. Efstathiou, Rowan-Robinson \& Siebenmorgen 2000). 

\acknowledgments The authors wish to thank Eli Dwek for providing the DIRBE and FIRAS data
used in Figure \ref{fig:dust_spectrum_isrf} in a machine readable form and for helpful
discussions.  Comments from the referee Dr. Bruce Draine improved the presentation and
clarity. This work has been partially supported by NASA grants NAG5-9203 and NAG5-7933.
KAM gratefully acknowledges financial support from the Louisiana Space Consortium through
NASA grant NGT5-40035 and the National Research Council through the Resident Research
Associateship Program. This research has made use of the NASA/IPAC Extragalactic Database
(NED; http://nedwww.ipac.caltech.edu).

\end{document}